\documentclass[aps,prm,amsmath,amssymb,reprint,superscriptaddress]{revtex4-2}

\usepackage{amsmath,charter,bm,graphicx,multirow}
\usepackage{graphicx}
\usepackage{dcolumn}
\usepackage{bm}
\usepackage{float}
\usepackage{color,soul}
\usepackage[english]{babel}
\usepackage[autostyle]{csquotes}
\usepackage{hyperref}

\begin{document}
	
	\title{\texorpdfstring{Phase-Separated Charge Order and Twinning Across Length Scales in CsV$_3$Sb$_5$}{Phase-Separated Charge Order and Twinning Across Length Scales in CsV3Sb5}}
	
	
	\author{Jayden Plumb}
	\affiliation{Department of Mechanical Engineering, University of California, Santa Barbara, California 93106, USA}
	\altaffiliation{These authors contributed equally to this work.}
	
	\author{Andrea Capa Salinas}
	\affiliation{Materials Department, University of California, Santa Barbara, California 93106, USA}
	\altaffiliation{These authors contributed equally to this work.}
	
	\author{Krishnanand Mallayya}
	\affiliation{Department of Physics, Cornell University, Ithaca, NY, 14853, United States}
	
	\author{Elliot Kisiel}
	\affiliation{Physics Department, University of California, San Diego, California 92093, USA}
	\affiliation{Advanced Photon Source, Argonne National Laboratory, Argonne, IL, 60439, USA}
	
	\author{Fellipe B. Carneiro}
	\affiliation{Centro Brasileiro de Pesquisas Físicas, 22290-180, Rio de Janeiro, RJ, Brazil}
	
	\author{Reina~Gomez}
	\affiliation{Materials Department, University of California, Santa Barbara, California 93106, USA}
	
	\author{Ganesh~Pokharel}
	\affiliation{Materials Department, University of California, Santa Barbara, California 93106, USA}
	
	\author{Eun-Ah Kim}
	\affiliation{Department of Physics, Ewha Womans University, Seoul 03760, South Korea}
	
	\author{Suchismita Sarker}
	\affiliation{Cornell High Energy Synchrotron Source, Cornell University, Ithaca, New York 14853}
	
	\author{Zahirul Islam}
	\affiliation{Advanced Photon Source, Argonne National Laboratory, Argonne, IL, 60439, USA}
	
	\author{Sam Daly}
	\affiliation{Department of Mechanical Engineering, University of California, Santa Barbara, California 93106, USA}
	
	\author{Stephen D. Wilson}
	\affiliation{Materials Department, University of California, Santa Barbara, California 93106, USA}
	\email[email: ]{stephendwilson@ucsb.edu}

	\begin{abstract}
		
		We present X-ray scattering studies resolving structural twinning and phase separation in the charge density wave (CDW) state of the kagome superconductor CsV$_3$Sb$_5$. The three-dimensional CDW state in CsV$_3$Sb$_5$ is reported to form a complex superposition of Star of David (SoD) or Tri-Hexagonal (TrH) patterns of distortion within its kagome planes, but the out-of-plane stacking is marked by metastability. In order to resolve the impact of this metastability, we present reciprocal space mapping and real-space images of CsV$_3$Sb$_5$ collected across multiple length scales using temperature-dependent high-dynamic range mapping (HDRM) and dark-field X-ray microscopy (DFXM). The experimental data provide evidence for a rich microstructure that forms in the CDW state. Data evidence metastability in the formation of $2\times 2\times 4$ and $2\times 2\times 2$ CDW supercells dependent on thermal history and mechanical deformation. We further directly resolve the real space phase segregation of both supercells as well as a real-space, structural twinning driven by the broken rotational symmetry of the CDW state. Our combined results provide insights into the role of microstructure and twinning in experiments probing the electronic properties of CsV$_3$Sb$_5$ where rotational symmetry is broken by the three-dimensional charge density wave order but locally preserved for any single kagome layer.
		
	\end{abstract}
	
	\maketitle

	\section{Introduction}
\begin{figure}[t]
		\centering
		\includegraphics[width=1.0\columnwidth]{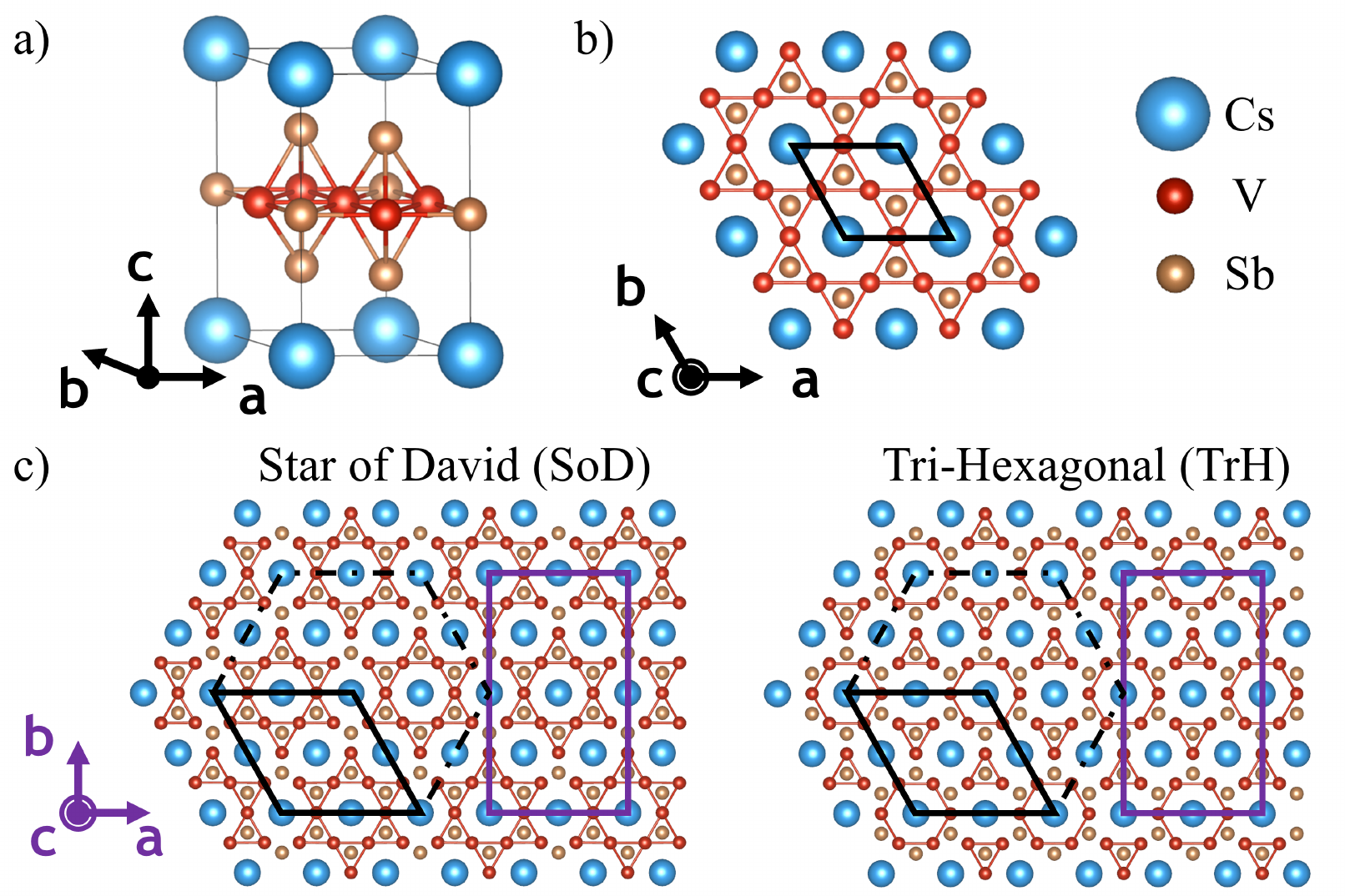}
		\caption{(a) View of the undistorted unit-cell of CsV$_3$Sb$_5$ at room temperature. (b) Crystal structure projection of the ab-plane, with V-V bonds drawn to illustrate the kagome network of vanadium atoms. (c) Crystal structure of the two distortion modes, Star of David (SoD) and its inverse Tri-Hexagonal (TrH), which occur below 94 K \cite{kautzsch2023structural}. Black lines indicate the nominal hexagonal structure of each state and purple lines indicate the orthorhombic structure formed below T$_{CDW}$.}
		\label{fig:CVScell}
	\end{figure}

	The $A$V$_3$Sb$_5$ ($A$= K, Rb, Cs) family of compounds are comprised of quasi-2D planes composed of kagome networks of vanadium ions coordinated by antimony and separated by planes of alkali metal ions (see Fig. \ref{fig:CVScell}(a,b)) \cite{ortiz2019new}. All three compounds in the material family host a primary charge density wave (CDW) instability, inside of which a number of unconventional phenomena are reported. These include a giant anomalous Hall effect \cite{yang2020giant, yu2021concurrence}, various hints of staged order \cite{zhao2021cascade, li2023unidirectional, luo2022possible, Wilson2024}, broken time-reversal symmetry \cite{jiang2021unconventional,mielke2022time, xing2024optical,le2024superconducting, xu2022three}, and unconventional superconductivity \cite{ortiz2020cs, ortiz2021superconductivity, yin2021superconductivity}. The most complicated evolution of electronic states within the CDW phase is reported to occur within the $A$=Cs member, CsV$_3$Sb$_5$, which also possesses the most complex parent CDW state \cite{Wilson2024}. 
	
	Specifically, both KV$_3$Sb$_5$ and RbV$_3$Sb$_5$ exhibit a common CDW state characterized by a 3\textbf{q}-type distortion of the kagome planes into tri-hexagonal (TrH) layers that stagger by half of a unit cell when stacked along the interplanar direction, creating a $2\times 2\times 2$ supercell \cite{frassineti2023microscopic, kautzsch2023structural}. This three-dimensional cell reduces the in-plane rotational symmetry of the unit cell to two-fold while largely preserving the six-fold rotational symmetry of each individual kagome plane. In contrast, CsV$_3$Sb$_5$, in its average structure, exhibits signs of both TrH- and star of David (SoD)-type breathing modes within its kagome network \cite{kautzsch2023structural, kang2023charge, hu2022coexistence}. The stacking of these distortions along the $c$-axis creates a larger $2\times 2\times 4$ supercell that also reduces the overall rotational symmetry to two-fold. Additionally, this larger cell seemingly competes with a smaller $2\times 2\times 2$ cell, and reports of their interplay differ \cite{stahl2022temperature, xiao2023coexistence}.
	
	While both angle-resolved photoemission and X-ray diffraction studies report the presence of both SoD and TrH distortions in CsV$_3$Sb$_5$ \cite{kautzsch2023structural, kang2023charge, hu2022coexistence}, X-ray scattering reports differ on the presence and thermal evolution of the $c$-axis modulation of the CDW state. Initial reports included observations of either a  $2\times 2\times 2$ cell \cite{li2021observation} or a larger $2\times 2\times 4$ superlattice \cite{ortiz2021fermi}. Studies of the thermal evolution of CDW order later showed that both states manifest in the same crystal with either the $2\times 2\times 4$ state \cite{stahl2022temperature} or the $2\times 2\times 2$ state \cite{xiao2023coexistence} stabilized via thermal quenching. This complexity and variance between samples and experiments suggests a shallow energy landscape with nearly degenerate patterns of CDW ordering \cite{wu2022charge}.
	
\begin{figure}[!ht]
		\centering
		\includegraphics[width=1.0\columnwidth]{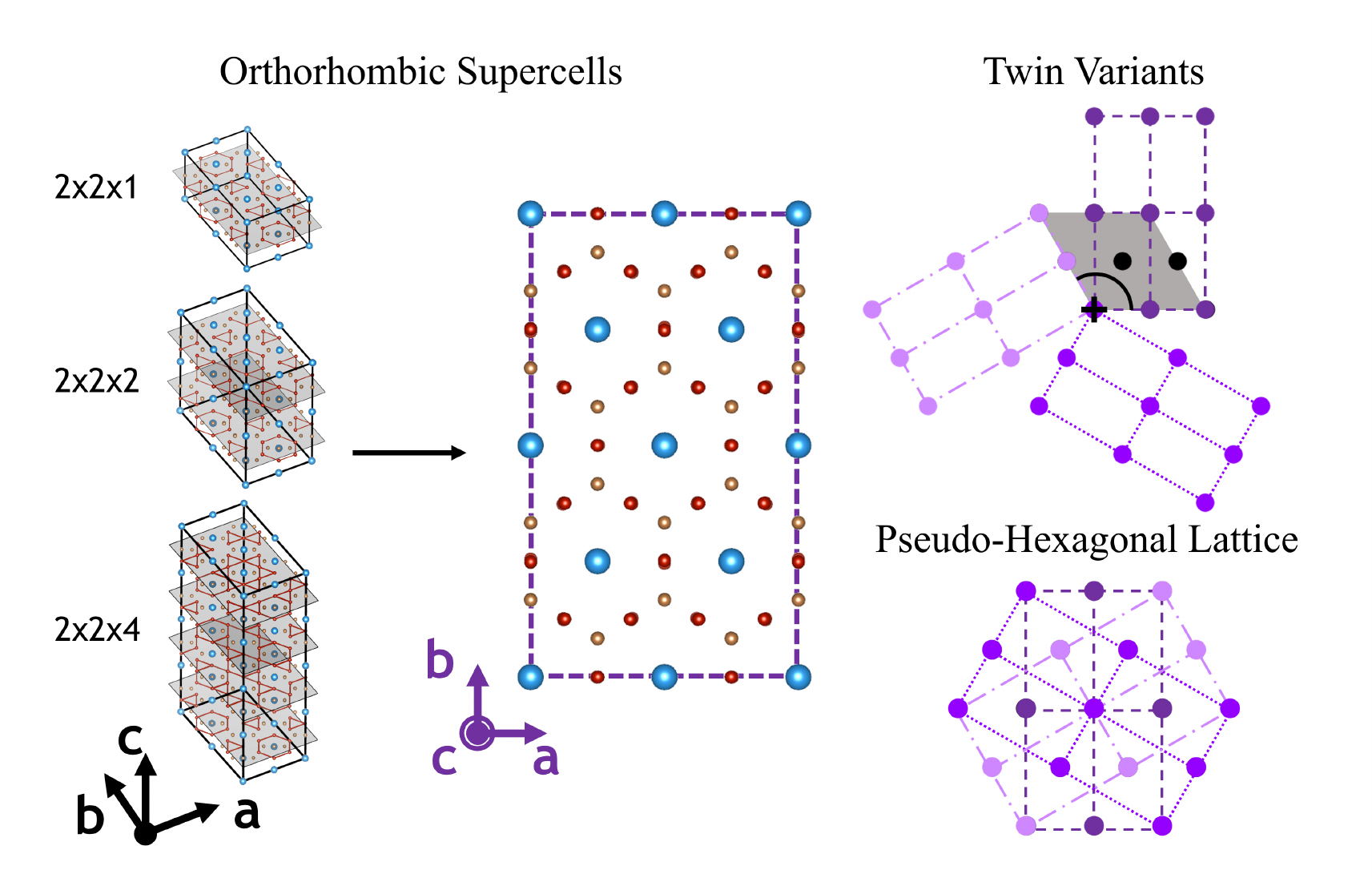}
		\caption{(left) Diagram showing the possible low temperature super-structures of CsV$_3$Sb$_5$, including the orthorhombic $2\times 2\times 1$, $2\times 2\times 2$, and $2\times 2\times 4$ cells respectively. A projection of the orthorhombic $ab$-plane is also displayed. (top right) Schematic of real-space hexagonal lattice points (black) overlaid with the three orthorhombic twin variants (purples), aligned by their respective origins. (bottom right) Tri-twinned orthorhombic unit-cells, shifted to be aligned by their respective centers, displayed to demonstrate how local two-fold symmetry, can combine into a pseudo-hexagonal lattice with apparent six-fold symmetry across the average structure. The three twin variants are rotated by 120$^\circ$ about the c-plane normal and are displayed using varied shades of purple and line dash styles.}
		\label{fig:supercell_twinning}
	\end{figure}

	Upon cooling below $T_{\mathrm{CDW}}$ = 94 K in CsV$_3$Sb$_5$, metastability and staging of multiple patterns of CDW ordering were reported in recent X-ray scattering measurements \cite{kautzsch2023structural}. Order along three different wave vectors was observed with differing onset temperatures: a $2\times 2\times 1$ structure forming just below the transition temperature followed by a $2\times 2\times 2$ structure forming below 90 K and lastly a $2\times 2\times 4$ structure that forms below 85 K (See Fig. \ref{fig:supercell_twinning}). Interestingly, the $2\times 2\times 1$ structure diminishes at the onset of the $2\times 2\times 4$ structure's nucleation, suggesting competition between stacking orders. Initial X-ray investigations of potential CDW phase separation reported domain formation on the scale of hundreds of microns \cite{xiao2023coexistence}, consistent with optical studies \cite{xu2022three} and scanning tunneling microscopy (STM) \cite{li2022rotation}. How these various CDW states nucleate and their interplay with the underlying microstructure of the crystal remains an active area of investigation. 

	The structural twinning below $T_{\mathrm{CDW}}$ seen in optical birefringence measurements \cite{xu2022three}, is consistent with the existence of three orthorhombic twin domains rotated by 120$^\circ$ from one another. The diagram in Fig. \ref{fig:supercell_twinning} illustrates how the rotational-twins, each with two-fold symmetry, can be combined to reproduce a pseudo-six-fold symmetry. This complicates the interpretation of conventional X-ray scattering data, and a multi-length scale approach that combines momentum-resolved probes with spatial resolution below several hundred microns has yet to be employed to study CsV$_3$Sb$_5$. This is relevant as understanding how these structural textures manifest and their intrinsic length scales remains vital to modeling the various symmetries reportedly broken below $T_{\mathrm{CDW}}$ in CsV$_3$Sb$_5$ \cite{Wilson2024}. 
	
	In this paper, we report a study of the CDW metastability and phase separation as well as twin domain formation in CsV$_3$Sb$_5$. A multi-length scale approach is used, by first reporting the impact of thermal history and crystallinity on CDW order using high-dynamic range X-ray mapping (HDRM). These measurements are performed using an $\approx300\times 800$ $\mu m^2$ X-ray footprint that averages across a rotated crystal using a conventional transmission setup. The resulting macroscopic average resolves the thermal evolution of $2\times 2 \times 2$ and $2\times 2 \times 4$ superstructures when the CDW is entered via slow versus fast rates of cooling and different CDW clusters are identified with differing onset temperatures. These measurements also reveal the impact of mechanical deformation in suppressing the $2\times 2 \times 4$ CDW state. 
	
	Complementary low-temperature, dark-field X-ray microscopy (DFXM) measurements are also reported across a textured crystal of CsV$_3$Sb$_5$ to interrogate the length scale of heterogeneity in CsV$_3$Sb$_5$'s CDW and crystallographic domain structures. Structural domain formation and phase separation between CDW states was observed with evidence of CDW domain sizes on the order of 100s of $\mu m$. Our results stress the importance of accounting for metastability, phase separation, and structural texture in studies of the electronic properties of CsV$_3$Sb$_5$.
	
	\begin{figure*}[t]
		\centering
		\includegraphics[width=1.0\textwidth]{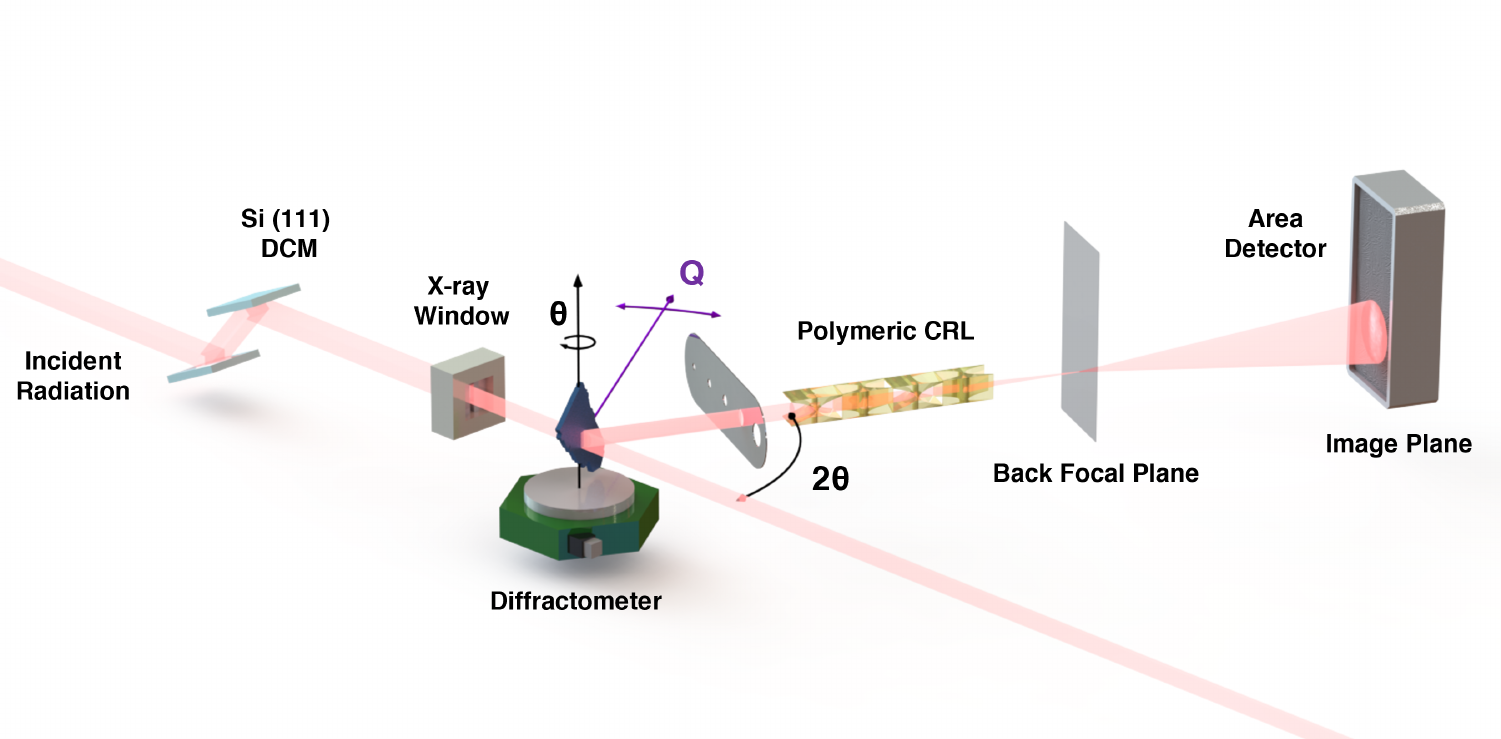}
		\caption{Schematic of the dark-field X-ray microscopy (DFXM) experiment configuration. Major components of the setup are labeled, including a diffractometer, X-ray objective lens, and an area detector placed at the image plane. The X-ray path is visualized using a light red beam. The vertical rotation axis, about which the sample is rotated in $\theta$ for mosaic determination, is denoted in black, while the momentum transfer \textbf{Q} is denoted in purple. Dimensions are not drawn to scale. }
		\label{fig:dfxmSchematic}
	\end{figure*}
	
	\section{Experimental Methods}
	\subsection{Sample Preparation}
	Plate-like crystals of CsV$_3$Sb$_5$ were grown using a crucible-based flux method detailed elsewhere \cite{ortiz2021fermi}. For the HDRM X-ray scattering measurements at CHESS, two high-quality samples from the same batch were analyzed via magnetic susceptibility measurements and exhibited the expected $T_{\mathrm{CDW}}=94$ K and bulk $T_{\mathrm{c}}=2.5$ K transitions. The first crystal studied via HDRM, `Crystal 1' was a bulk as-grown single crystal and the second crystal `Crystal 2' was a square cut out of a larger single crystal using a razor blade to be one-third its original length in \textit{a} and \textit{b} while maintaining its original thickness. The crystal studied in DFXM measurements was a separate, textured crystal that was mechanically exfoliated using tape to remove surface impurities and achieve a desired thickness of $\approx 100$ $\mu m$. This allowed for a transmission scattering geometry. 
	
	\subsection{HDRM X-ray measurements} \label{HDRM}
	High-dynamic range X-ray scattering measurements were performed at the CHESS QM2 beam line using an incident wavelength of $\lambda=0.47686$~\AA. A helium and nitrogen switchable-option blower was used to cool crystals down to 35 K. Cooling was driven via nitrogen flow down to 100 K and then switched to He-gas flow below this temperature. ``Slow-cooled" samples were treated by cooling a crystal from room temperature to 150 K at 5~K/min, then cooling from 150 K to 100 K at 2~K/min, followed by cooling at 1~K/min through the CDW transition down to 35~K. ``Fast-cooled" samples were  cooled at 10~K/min from 300~K to 35~K. Data were collected in transmission on a 6-megapixel detector by performing three-axis 360$^{\circ}$ rotation scans at 0.1$^{\circ}$ step size during warming. The NeXpy software package \cite{osbornenexpy} was employed to visualize and perform one-dimensional cuts of reciprocal space maps. APEX3 software \cite{APEX3} was used to solve the low-temperature structure.
	
	An unsupervised machine learning tool called X-ray Temperature Clustering (X-TEC) was used for analysis of the parametric data sets. X-TEC uses a Gaussian Mixture Model to identify peaks in reciprocal space with distinct temperature dependencies. In order to distinguish the functional form  of the intensity-temperature trajectory rather than their magnitudes, the intensity of each momenta $\vec{q}$ is re-scaled (z-scored) as $\tilde{I}_{\vec{q}}(T)=\left(I_{\vec{q}}(T)-\mu_{\vec{q}}\right)/\sigma_{\vec{q}}$ where $\mu_{\vec{q}}$ is the mean over temperature $T$, and $\sigma_{\vec{q}}$ is the standard deviation in $T$ \cite{Venderley2022Proc.Natl.Acad.Sci.}.
	
	\subsection{DFXM Experimental Setup} \label{dfxmsetup}
	DFXM is an augmented form of single-crystal X-ray diffraction (XRD) in which an X-ray objective is placed in the path of the diffracted beam to spatially distinguish and magnify the origins of the diffracted signal. Real-space images that originate from a specific crystallographic Bragg peak can subsequently be collected on a detector placed at the image plane. A combination of high spatial (real-space) and angular (reciprocal-space) resolutions make DFXM an ideal technique for multi-scale investigations \cite{simons2016multiscale, yildirim2020probing}. This technique, for instance, has previously been used to study a low temperature transition in a layered oxide, NaMnO$_2$, textured across multiple length scales\cite{plumb2023dark}. 
	
	DFXM measurements were performed at the 6-ID-C beam line of the Advanced Photon Source (APS) at Argonne National Laboratory. A generalized schematic of the DFXM setup, not including the low temperature cryostat, is displayed in Fig. \ref{fig:dfxmSchematic}. The experimental setup at Sector 6 nominally incorporates the following, listed in order (starting from the undulator/source) along the incident X-ray direction: a Si (111) double-crystal monochromator, tuned to transmit between \(20\) and \(22\) keV X-rays; a Huber X-ray slit screen that controlled the incident spot size on the sample; an X-ray-diffraction capable cryostat \cite{plumb2023dark}; a 50 $\mu m$ pinhole placed just in front of the objective lens to reduce noise from the divergence of the diffracted signal; a polymeric compound refractive lens (CRL), designed for \(20\) keV X-rays, used as the objective lens with a focal length of 131 $mm$ and working distance of 140 $mm$ \cite{opolka2021multi, qiao2020large}; and two area detectors mounted on a translation/rotation stage, to capture the magnified diffraction signal at the image plane. The diffracted illumination was ultimately limited to 20 $\mu$m in width by the incident slit screen and 50 $\mu$m in height by the exit pinhole.
	
	The inherent X-ray absorption of the DFXM technique, presents a significant challenge to imaging weak diffraction peaks, such as those formed by charge density waves. To overcome this challenge, a large-area Medipix3-based-detector array (LAMBDA) from X-Spectrum was used to image the CDW signal at $\textbf{q}=(\frac{1}{2}, \frac{1}{2}, \frac{1}{2})$ and $\textbf{q}=(\frac{1}{2}, \frac{1}{2}, \frac{1}{4})$. This direct X-ray area detector hosts a native pixel size of \(55\) microns. CDW images collected using this detector, paired with the X-ray magnification of the CRL, yield an effective pixel size of 2.1 $\mu m$. A second, higher resolution area detector, was employed to capture images of CsV$_3$Sb$_5$'s brighter, primary structural peaks. This detector is composed of a 20 $\mu m$ LuAG:Ce X-ray scintillator, reflective mirror, $5\times$ optical objective lens, and a 5.5-megapixel sCMOS Zyla camera made by Andor. With a native pixel size of 6.5 $\mu m$, the effective pixel size of images collected using the Zyla detector is 0.05 $\mu m$. The resulting experimental feature resolution, or the smallest resolved line-pair, of the LAMBDA and Zyla detectors are $\approx2~\mu m$ and $\approx400~nm$, respectively. In DXFM measurements, the crystal was initially oriented such that the $[0, 0, 1]$ direction and the reciprocal lattice vector $[1, 1, 0]$ were parallel and perpendicular to the incident beam respectively.
	
	
	\subsection{DFXM Data Analysis} \label{dfxmdata}
	All DFXM data were collected using rocking curve imaging (RCI), whereby multiple images are collected at a single Bragg peak as the sample is rotated through various orientations ($\theta$). Rotating the sample in this way provides a method for sub-domain discrimination as the diffracted illumination only appears when the associated material satisfies the Bragg condition. The resulting data is a series of images that correspond to the various sample orientations for a given sample position and diffraction angle. The small numerical aperture of the objective lens and the large sample-to-detector distance ($\approx$ 2 $m$) lead to an exceptional angular resolution of 0.001$^{\circ}$. Integrating the intensity across DFXM images and plotting it against $\theta$ produces a one-dimensional curve, emblematic of point detector intensity curves generated from conventional X-ray diffraction. Applying this concept on a pixel-by-pixel basis, assuming a 1:1 mapping of pixels to material points, results in a spatially resolved plot of rocking curve information. 
	
	Due to a small numerical aperture, axial (depth) resolution along the diffracted beam direction (i.e. $k^\prime$) is poor. In the current setup, a transmission horizontal scattering geometry was used. So, the depth resolution is given to be an effective width of the incident beam along an intersection of a line parallel to $k^\prime$. As a consequence, it is not possible to isolate CDW or twin domains lying along $k^\prime$.  While tomographic approaches may be used to obtain depth sensitivity, it requires a complete azimuthal rotation around the scattering vector, which was not possible given mechanical restrictions with a cryostat and sample morphology. Nevertheless, a 50 $\mu$m tall horizontally, 20 $\mu$m wide beam was used to illuminate a thin slice of the crystal to reduce the number of domains contributing to a given pixel in each image, which was obtained by translating the sample through the incident beam. As a result, clean domain boundaries were identified in DFXM images. In the future, the use of coded aperture to generate structured illumination can be used for a complete 3D reconstruction of CDW domain network as was demonstrated in a recent study of nematic order in a pnictide superconductor \cite{gursoy2024dark}.
	
	Background subtraction based on the work of \citet{garriga2023darfix} was performed on all collected images to enhance contrast before analysis. Subtractions employed a combination of up to three major steps, including dark-detector subtraction, hot pixel removal, and thresholding. After noise removal was completed, images were analyzed by calculating key spectral parameters, characteristic of the periodic structure of the system. The analysis followed a procedure whereby the intensity of each pixel was plotted across $\theta$ values (slices) and fit with a Gaussian curve. The parameters extracted from this fitting were the maximum intensity or amplitude of the peak, the peak's center of mass ($\mu$) or mosaic, the full-width at half maximum (FWHM) of the curve, and the integrated area under the curve. These parameters are plotted across every pixel in a representative image, producing visualizations that highlight contrasts that stem from different material domains/defects. Examples of these plots are displayed and discussed further in later sections of this manuscript. 
	
	\section{Results and Discussion}
	
	\subsection{HDRM Temperature Studies: Strain Dependence and Cooling Hysteresis}
	
	\begin{figure}[t]
		\centering
		\includegraphics[width=1.0\columnwidth]{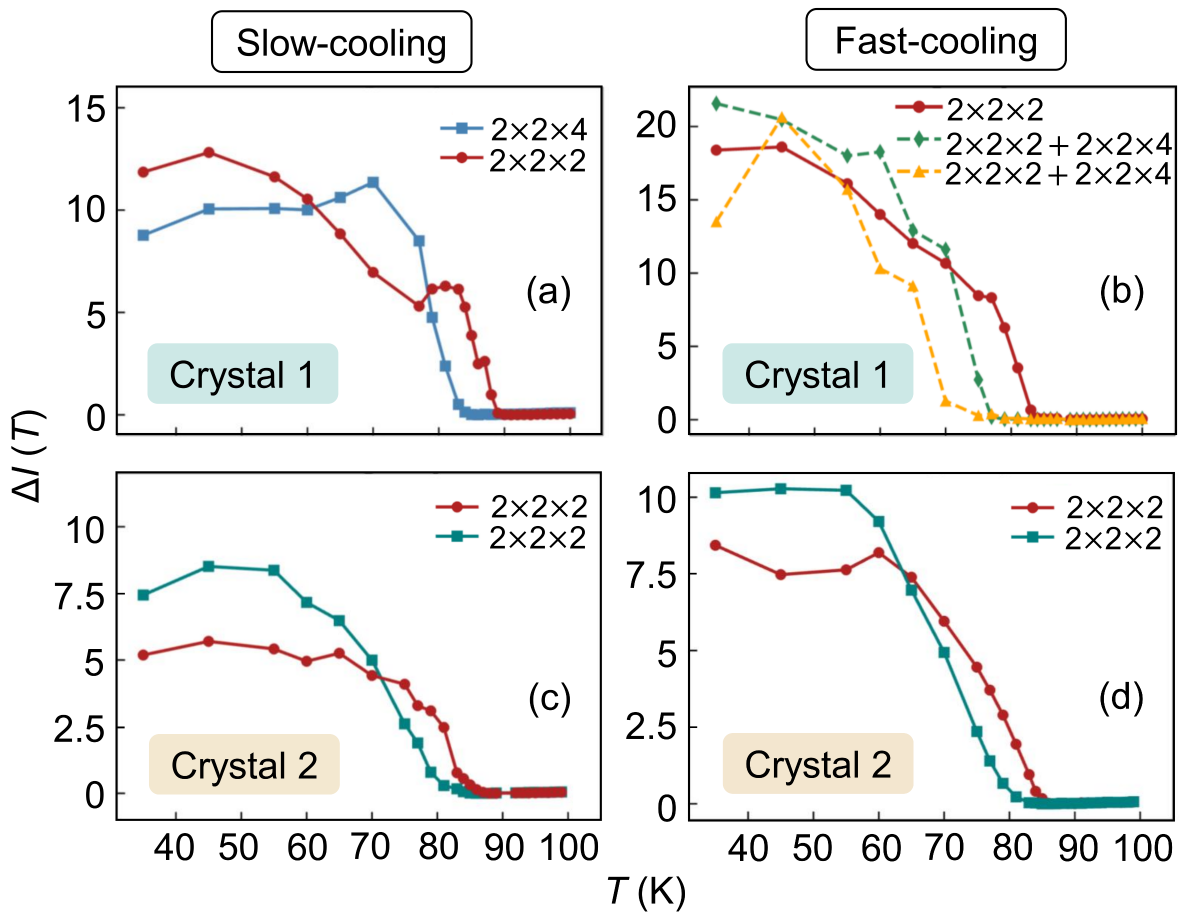}
		\caption{X-TEC analysis revealing the distinct intensity-temperature trajectories for (a), (b) 'Crystal 1'  or as-grown crystal, and (c), (d) 'Crystal 2'  or cut crystal. Both crystals underwent identical slow cooling (a, c) and fast cooling (b, d) protocols. Plots show cluster-averaged intensities $\Delta I(T)$ where the minimum value of the average trajectory is subtracted.
			Further, a quarter-type peak and a half-type peak whose pixel intensity tracks the same temperature-dependent trajectory, will be classified under one of the two mixed $2\times 2\times 2$ + $2\times 2\times 4$ clusters.}
		\label{fig:XTEC}
	\end{figure} 
	
	Two crystals were explored using X-ray diffraction under variable cooling rates. `Crystal 1', which is a pristine as-grown bulk crystal, and `Crystal 2', which is a subsection of an as-grown bulk crystal from the same batch that has been mechanically excised with a razor blade. Both crystals were subjected to identical slow-cooling and fast-cooling cycles, and the onset of CDW order was analyzed via the X-TEC algorithm. The results are plotted in Fig. \ref{fig:XTEC}.
	
	Focusing first on `Crystal 1,' two distinct clusters are identified upon warming after a ``slow cooling" protocol, as shown in Fig. \ref{fig:XTEC} (a). One of these corresponds to the $2\times 2\times 2$ CDW state and one to the $2\times 2\times 4$ CDW state. Upon warming, the $2\times 2\times 4$ state vanishes above 83 K, which corresponds to the end of the plateau in the thermal evolution of the $2\times 2\times 2$ cluster, and this is followed by the $2\times 2\times 2$ cluster vanishing above 88 K. This onset temperature for CDW order is lower than the nominal $T_{\mathrm{CDW}}=94$ K and likely arises from a thermometry offset between the cryostat and sample temperature. Magnetization measurements on this crystal show the expected transition temperature. Regardless of this offset, the thermal evolution of the two clusters suggests competition with one another, where the onset of one cluster impacts the relative intensity of the other.
	
	Exploring this further, the same `Crystal 1' was then studied on warming following a ``fast cooling" protocol, with the results shown in Fig. \ref{fig:XTEC} (b). Two qualitative changes appear. The first is that CDW order vanishes above 83 K; this is 5 K lower than the onset of CDW order after slow cooling. The second change is that $2\times 2\times 4$ correlations no longer have a distinct thermal evolution. They instead mix with $2\times 2\times 2$ correlations and vanish above $\approx 75$ K with two separate temperature-dependent order parameters identified. We note here that the mixed $2\times 2\times 2$ and $2\times 2\times 4$ clusters, shown in Fig. \ref{fig:XTEC}(b), are indistinguishable to the machine learning algorithm X-TEC employs because the algorithm classifies pixel intensity according to its temperature dependence, not $(H, K, L)$ position. This is consistent with a first-order phase boundary into an ordered phase with nearly degenerate CDW configurations, similar to results reported in Xiao et al \cite{xiao2023coexistence} with the added insight of distinct $2\times 2\times 2$ and $2\times 2\times 4$ reflection clusters mixing within a common onset temperature in the quenched state. 
	
	A second crystal grown from the same batch was then tested, `Crystal 2.' Instead of using an as-grown crystal, which is typically larger than that needed for an X-ray diffraction measurement, `Crystal 2' was obtained by using a razor blade to mechanically cut a smaller footprint out of a larger parent crystal. As crystals of $A$V$_3$Sb$_5$ behave as mechanically soft foils, this is a reasonably common procedure for certain measurements of CsV$_3$Sb$_5$ where the size needs to be reduced. The results of slow and fast cooling X-ray diffraction measurements on `Crystal 2' are shown in Fig. \ref{fig:XTEC} (c) and (d). 
	
	\begin{figure}[!ht]
		\centering
		\includegraphics[width=1.0\columnwidth]{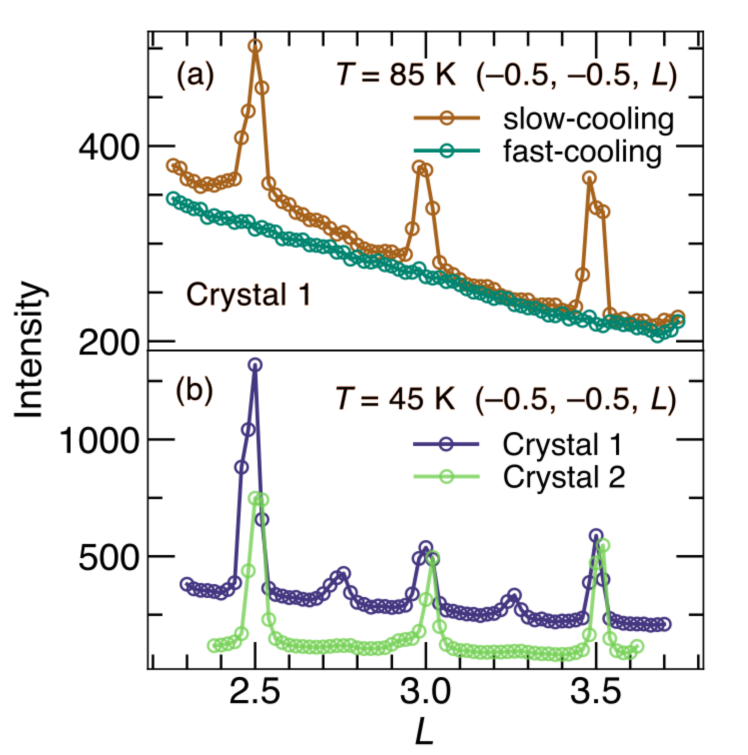}
		\caption{\textit{L}-cuts about $(H,K)=(-0.5, -0.5)$ showing: (a) CDW superlattice peaks in `Crystal 1' following slow cooling and fast cooling at 85~K, and (b) absence of quarter-type peaks in `Crystal 2' at 45~K.}
		\label{fig:Cuts}
	\end{figure}

	Strikingly, $2\times 2\times 4$ correlations are absent in this cut crystal. Instead, only two $2\times 2\times 2$ correlation clusters are identified with onset temperatures $\approx 5$ K offset from one another at 83 K and 88 K. These two clusters correspond to two domains within the sample, and each cluster maps to one crystallite or the other. Both fast cooling and slow cooling protocols yield qualitatively identical trajectories for the two CDW clusters, signifying that the metastability and thermal hysteresis has also vanished. Inspection of the primary Bragg peaks for this cut crystal shows a slight broadening relative to those in as-grown `Crystal 1', suggesting that subtle mechanical deformation induced during cutting is likely responsible for quenching the $2\times 2\times 4$ state.

	To fully illustrate some of these trends, representative regions of reciprocal space were chosen to show data following the conclusions drawn from the X-TEC analysis. An integrated cut along L at $H=-$0.5, $K =-$0.5 at 85~K is shown for `Crystal 1' in Fig. \ref{fig:Cuts} (a). The suppression of $T_{\mathrm{CDW}}$ under quenching determined from the cluster analysis is evident, and the slow cooled sample shows CDW order at this temperature while the fast cooled sample does not. A representative cut comparing the CDW order in `Crystal 2' and `Crystal 1' is shown in Fig. \ref{fig:Cuts}(b) where the $2\times 2\times 4$-type peaks are absent in Crystal 2. 
	
	The direct visualization of representative scattering data reinforces the conclusions drawn from the X-TEC-based analysis. Namely that the $2\times 2\times 4$ CDW state accompanies a metastability where close competition between CDW states with mixed wave vector and onset temperatures appears. Quenching through $T_{\mathrm{CDW}}$ modifies this phase competition and suppresses the ordering temperature. Upon introducing a slight mechanical deformation, evident via a slightly degraded crystallinity, the balance between these states is altered. With this small perturbation, the $2\times 2\times 4$ state is suppressed and the CDW state evolution becomes insensitive to quenching versus slow cooling through $T_{\mathrm{CDW}}$. A similar suppression of the $2\times 2\times 4$ state was reported due to slight alloying/disorder \cite{Kautzsch2023} and the balance between CDW states is also  tunable via external pressure \cite{Zheng2022}. 
		
	Due to time constraints associated with synchrotron beam time, this effect was demonstrated only on one sample. While there may be other factors beyond trivial strain that suppress the $2\times 2\times 4$ state, this should nevertheless stand as a warning that routine mechanical manipulation (e.g. cutting, polishing, gluing) of CsV$_3$Sb$_5$ in electronic properties measurements must be done carefully as to avoid changing the CDW state.  It is worthwhile to note the that impact of subtle mechanical perturbation on the electronic properties of CsV$_3$Sb$_5$ is also observed in strain-relieved magnetotransport studies\cite{guo2022switchable} and in the in-plane transport anisotropy \cite{guo2024correlated}.
	
	
	\begin{figure}[!ht]
		\centering
		\includegraphics[width=1.0\columnwidth]{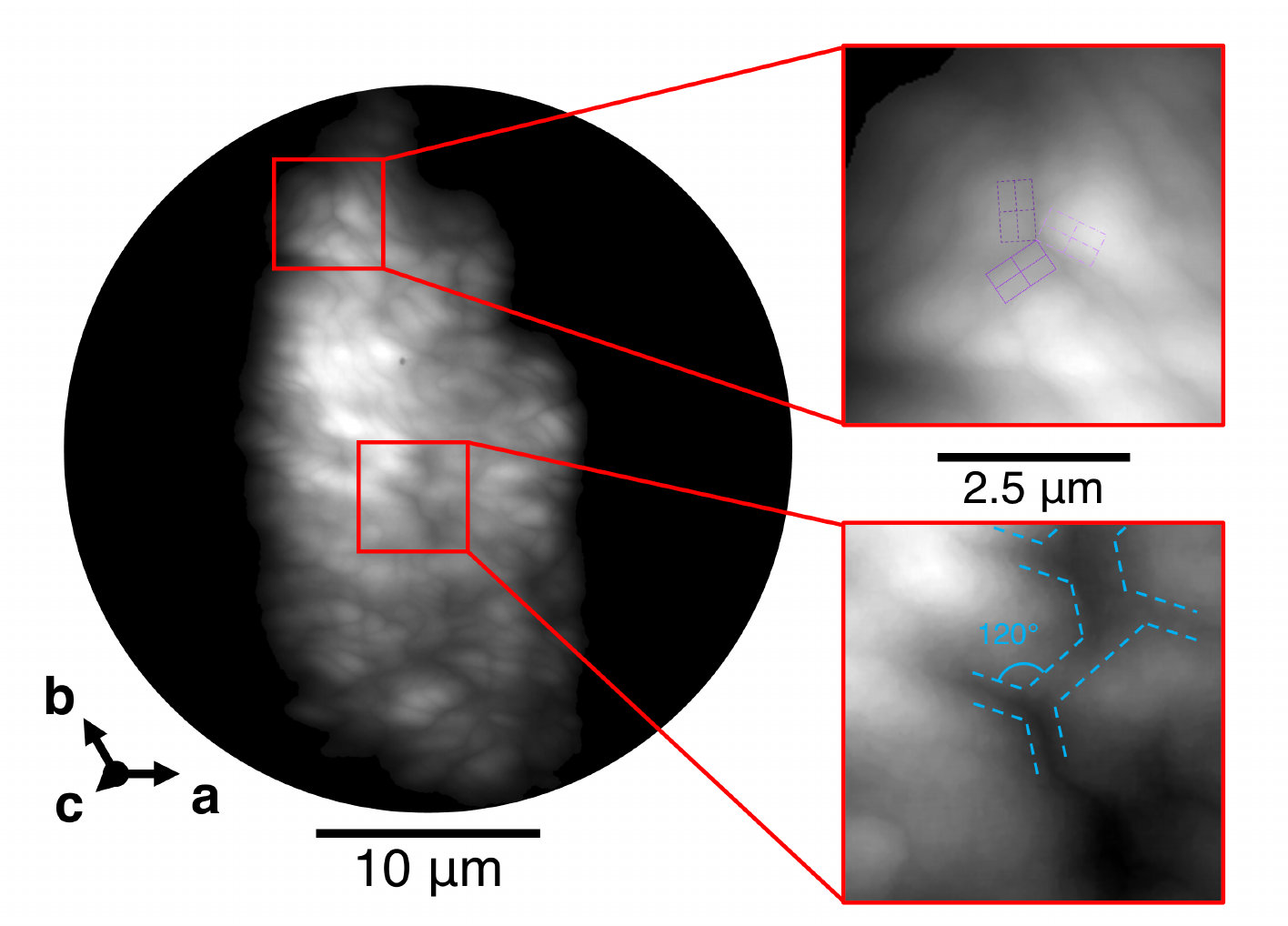}
		\caption{High-resolution image of the $(2, 2, 0)$ peak of CsV$_3$Sb$_5$ at low temperature (3.2 K), which illustrates a network of hundreds of crystal domains across the 32.5 $\mu m$ span of the image. Real-space crystal axes and a scale bar are included at the bottom of the image. Two cutouts of triple-point boundaries, outlined in red, have been magnified and presented to the right of the primary image. The top cutout is overlaid with the twinning diagram from Fig. \ref{fig:supercell_twinning} to highlight the experimental validation of the theoretical twinning model. The lower cutout is overlaid with cyan dashed lines that emphasize a network of triple-point boundaries that form larger hexagonal domains.}
		\label{fig:boundH}
	\end{figure} 
	
	\subsection{Microstructural Length Scales and Twin Formation}
	In order to gain deeper insights into the spatial impact of metastability in the CDW state of an as-grown crystal, DXFM measurements were performed. First, to resolve the CDW effect on the local microstructure of the $ab$-plane, a high-resolution RCI scan using the $(2, 2, 0)$ primary Bragg reflection $(2\theta = 23.61^\circ)$ was collected at $T = 3.2~\mathrm{K}$. The Zyla detector described in section \ref{dfxmsetup}, was employed to image this structural peak across $\theta$ positions in $0.001^\circ$ steps, with 5 seconds of exposure each. By evaluating the total intensity of each image with respect to the sample's theta position, an intensity curve akin to the results of single crystal diffraction can be obtained. For the $(2, 2, 0)$ peak, an average FWHM of $0.024^\circ$ is observed, which aligns with expectations for a highly crystalline sample. The scattering geometry of the $(2, 2, 0)$ peak means that the DFXM images are projections taken at a shallow angle $(13^\circ)$ from the $c$-axis, meaning any contrast observed will necessarily be an average representation of the structure along the interlayer stacking direction.

	
	\begin{figure*}[!t]
		\centering
		\includegraphics[width=1.0\textwidth]{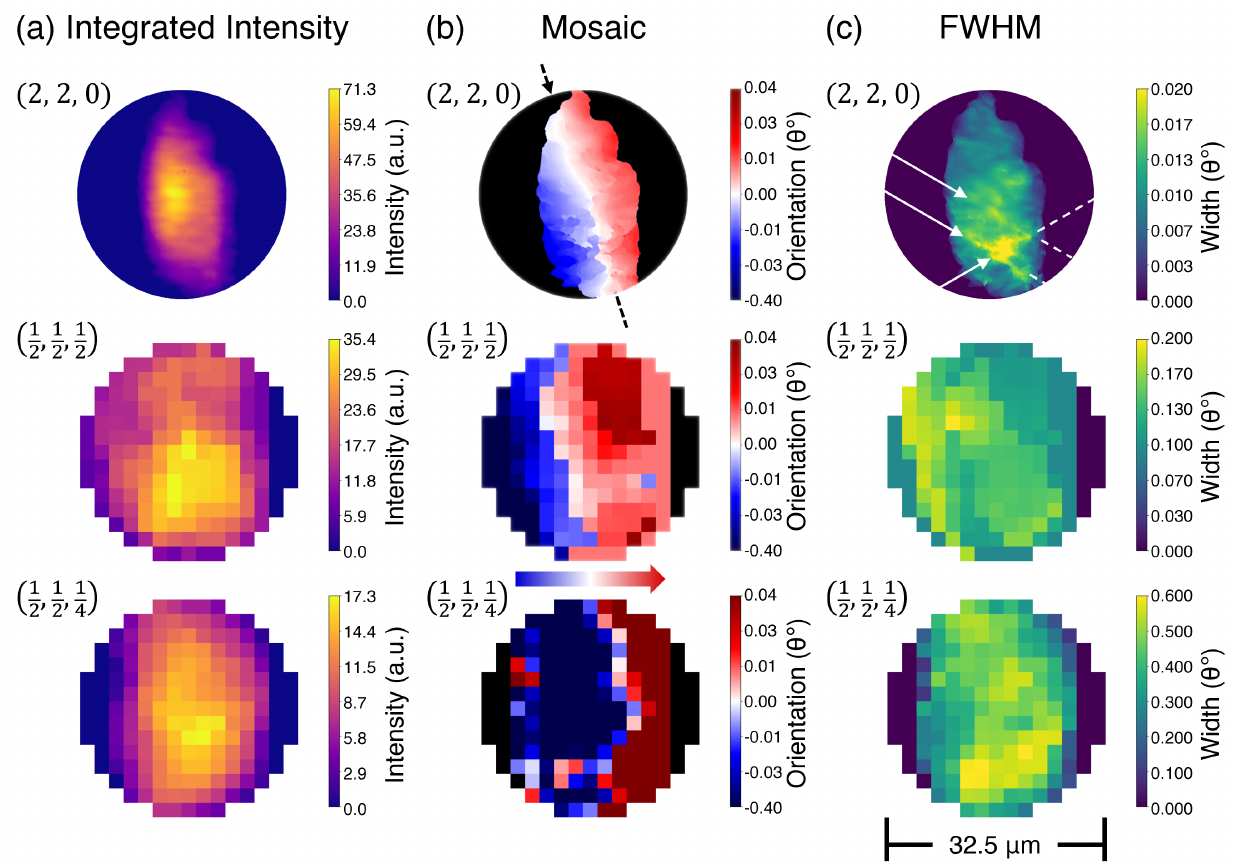}
		\caption{Overview and comparison of the DFXM data collected on CsV$_3$Sb$_5$, at a single position, from the (h, k, l) indicated in the top left corners. For each peak (h, k, l), the following is displayed from left to right. Plots of the integrated intensity of each pixel across theta positions are shown, characterizing the illuminated volumes of each peak's crystallographic domain of origin, projected to a 2D image. Plots of the relative misorientation (mosaic) of each pixel's fit center-of-mass compared to the image averaged center-of-mass are also presented. Mosaic plots highlight contrasts due to changes in crystal orientation and display mosaic boundaries. Lastly, plots of the full width at half maximum (FWHM) fit to each pixel are shown. FWHM values tend to increase with the concentration of local heterogeneities (\textit{i.e.} dislocations, faults, \textit{etc.}) within a given diffraction signal. White arrows and dotted lines have been added to the $(2, 2, 0)$ peak's FWHM plot to highlight $120^\circ$ separated boundaries, identified by their increased width values.}
		\label{fig:overview}
	\end{figure*}
	
	Plotting the maximum intensity of the $(2, 2, 0)$ images reveals a number of linear boundaries (see Fig. \ref{fig:boundH}) across the 32.5 $\mu m$ field-of-view. A number of triple-point boundaries appear that signify three in-plane 120$^\circ$ rotations. Two such regions, centered on boundaries with three-fold symmetry, have been highlighted and digitally magnified for presentation in Fig. \ref{fig:boundH}. These triple-point boundaries are likely coordination centers for rotational twinning born from lowering the original 6-fold hexagonal symmetry to 2-fold orthorhombic symmetry in the CDW state. These boundaries extend an average of $\approx3.7$ $\mu m$ from their centers, and faint two-dimensional modulations can be seen extending from these junctions with boundaries that are nearly perpendicular to its opposing triple-point boundary. The origin of these modulations is likely subsurface texture or terracing that modulates the intensity. 
	
	To explore this further, the integrated intensity, mosaic shift, and FWHM for the $(2, 2, 0)$ peak are plotted in Fig. \ref{fig:overview}. The mosaic plot of the $(2, 2, 0)$ peak shows a near-planar orientation gradient which reveals a transition between crystallographic domains. Notably, the central mosaic boundary follows a fault line that passes through multiple of the observed triple-point boundaries. This pattern enforces the idea of twin boundaries nucleating along larger (potentially pre-existing) hexagonal boundaries, with twin variants locally competing for formation. Additional texture is evident in the FWHM plot of the $(2, 2, 0)$ peak shown in Fig. \ref{fig:overview} where 120$^\circ$ rotated boundaries that extend over 10 $\mu m$ appear. This potentially arises from local strain fields from subsurface defects, though the effect is subtle in this field of view.
	

	\begin{figure*}[t]
		\centering
		\includegraphics[width=1.0\textwidth]{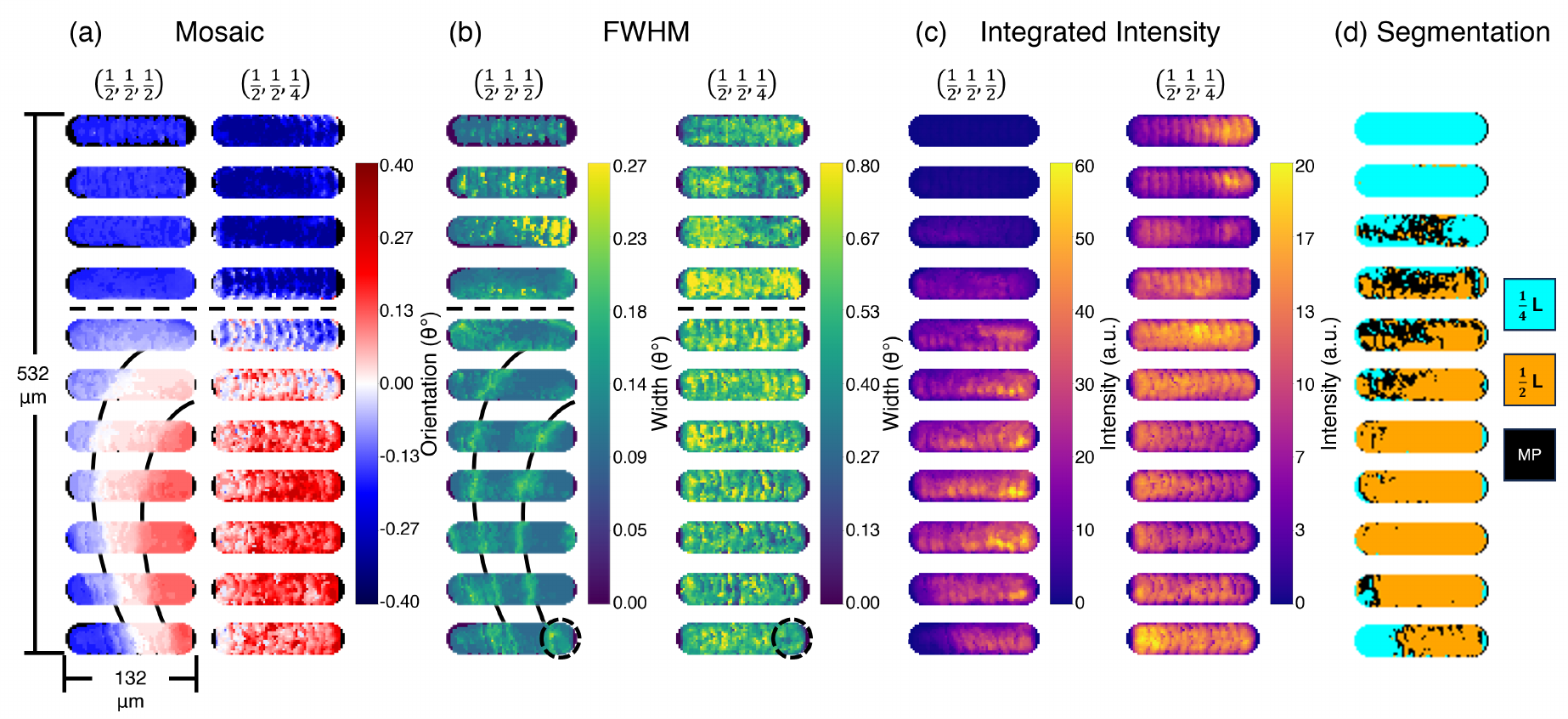}
		\caption{Plots of experimental parameters calculated from images stitched together based on sample position during area rastering. Parameters calculated from two charge density wave signals $(\frac{1}{2} L$ and $\frac{1}{4} L)$ are presented side-by-side for visual comparison. Displayed from left to right, are plots of the pixel-based mosaic spread, FWHM, and integrated intensity, shown with their respective color bars. Additionally, a corresponding segmentation plot is displayed, with each pixel denoting the dominate CDW signal (by integrated area) of its related material position. Black pixels in the segmentation plot denote areas in which there is a mixed peak \enquote{MP} signal, meaning that both the integrated intensity from each CDW signal is within 30\% of the other. Scale bars displayed on the leftmost plot indicate the total area spanned by the DFXM measurements. Black dashed circles in the bottom right of each FWHM plot denote the singular location of the plots presented in figure \ref{fig:overview}.}
		\label{fig:CDW}
	\end{figure*}
	
	\subsection{Charge Density Wave Phase Separation}
	At the same sample location, DFXM images were also collected at the wave vectors $\textbf{q}=(\frac{1}{2}, \frac{1}{2}, \frac{1}{2})$ and $\textbf{q}=(\frac{1}{2}, \frac{1}{2}, \frac{1}{4})$, corresponding to the $2\times 2\times 2$ and $2\times 2\times 4$ CDW states, respectively. The LAMBDA detector, described in section \ref{dfxmsetup}, was employed to image the weaker CDW peaks using a 10 second exposure time, and $0.1^\circ$ $\theta$-steps. One-dimensional intensity curves for both signals can be generated in a similar fashion to the $(2, 2, 0)$ peak for comparison. The maximum intensity observed for the $(\frac{1}{2}, \frac{1}{2}, \frac{1}{2})$ peak is 21,000 counts, while the maximum intensity for the $(\frac{1}{2}, \frac{1}{2}, \frac{1}{4})$ peak is 3,600 counts. The factor of 6 reduction in the peak intensity of the $(\frac{1}{2}, \frac{1}{2}, \frac{1}{4})$ signal coincides with a factor of 3 increase in its FWHM $(0.472^{\circ})$, resulting in a factor of 2 reduction of the $\theta$-integrated intensity. 
	
	
	The integrated intensity, relative mosaic shift, and FWHM values plotted across the X-ray  beam footprint are again displayed in Fig. \ref{fig:overview}. Mosaic plots of the CDW peaks show a left-to-right orientation gradient for the $(\frac{1}{2}, \frac{1}{2}, \frac{1}{2})$ and $(\frac{1}{2}, \frac{1}{2}, \frac{1}{4})$ peaks, qualitatively similar to the mosaic boundary seen in the primary $(2, 2, 0)$ peak. The intensity distribution for each CDW peak is also qualitatively similar across the illuminated area, while the FWHM distributions of each peak show some subtle variance; however these maps suffer from the poorer resolution of the detector setup needed to resolve the CDW order. To explore further and at length scales outside of the 32.5 $\mu m$ field-of-view, the incident X-ray illumination was rastered across the sample.
	
	
	
	DXFM data were collected via a series of RCI scans that were performed for each peak at various sample locations. Images were taken at positions spanning \(500\) $\mu m$ in the vertical-direction and \(100\) $\mu m$ in the horizontal-direction, where images from a given peak were then stitched together according to each scan's relative position. The sample was moved between scans such that there was no overlap in the vertical-direction (50 $\mu m$ steps) and a 66\% overlap in the horizontal-direction (10 $\mu m$ steps). Fig. \ref{fig:CDW} shows the resulting parameter plots of the stitched DFXM images for the $(\frac{1}{2}, \frac{1}{2}, \frac{1}{2})$ and $(\frac{1}{2}, \frac{1}{2}, \frac{1}{4})$ peaks. The reduction in peak intensity and increase in the FWHM of the $(\frac{1}{2}, \frac{1}{2}, \frac{1}{4})$ peak persists at this larger length scale, and evaluating this larger surveyed area gives a more complete picture of the CDW texture in the crystal.
	
	
	\begin{figure*}[t]
		\centering
		\includegraphics[width=1.0\textwidth]{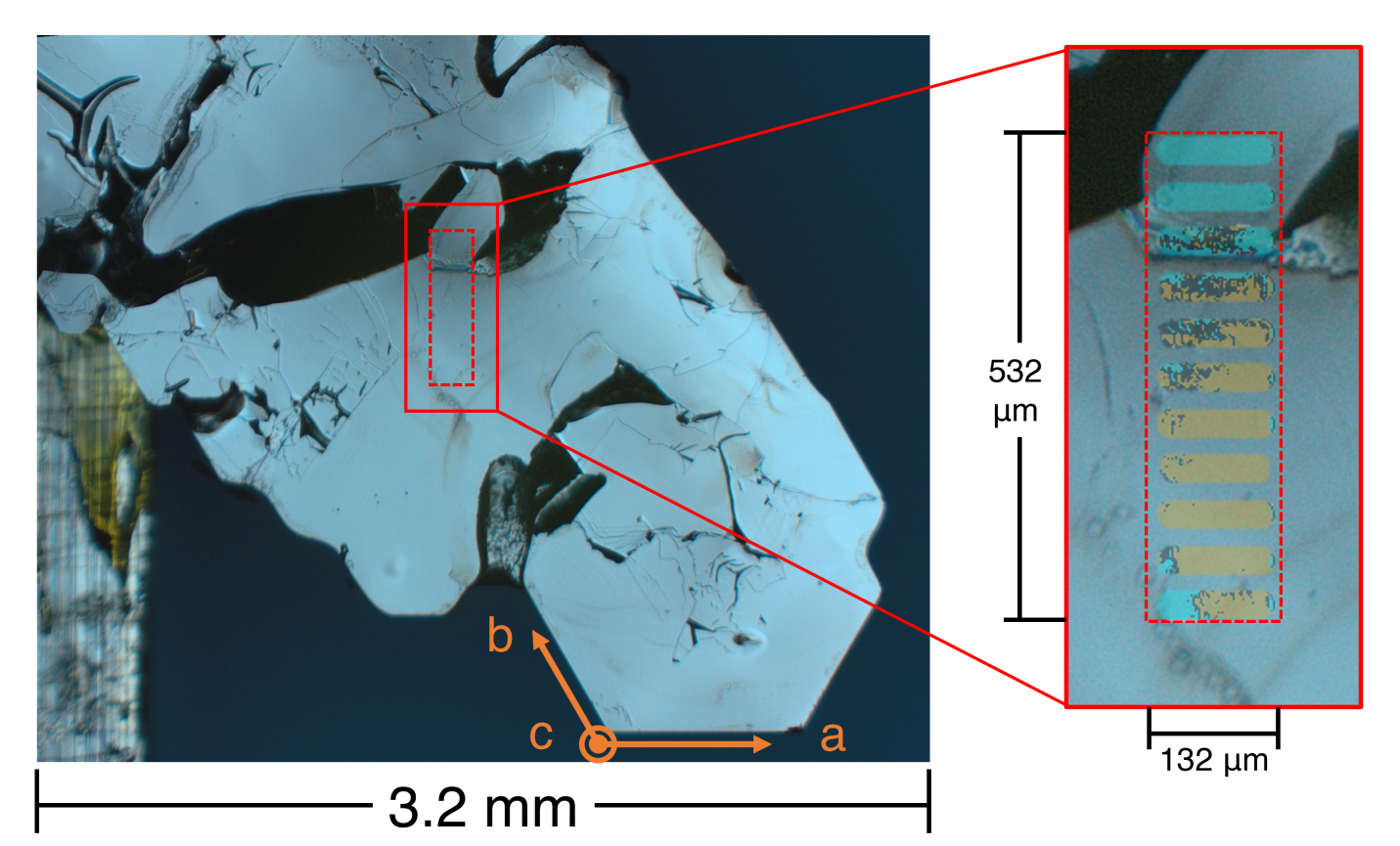}
		\caption{Optical microscope image of CsV$_3$Sb$_5$ sample, as it was measured in the DFXM experiment. Orange arrows indicate the real-space crystallographic orientation of the sample. Rectangles drawn on the image using solid and dashed red lines indicate the area of the magnified image cutout presented to the right and the relative area covered in the DFXM scans of the CDW peaks, respectively. The magnified cutout is displayed with an overlay of a semi-transparent version of the segmentation plot shown in figure \ref{fig:CDW}. This overlay highlights the relationship between the macroscale characteristics observed in the optical images with the mesoscale characteristics observed in the stitched DFXM images.}
		\label{fig:ScanVol}
	\end{figure*}
	
	Looking first at the mosaic shift maps, three distinct regions appear, colored to represent their relative negative and positive shifts from center respectively. The predominant mosaic boundary separates the scanned region into upper and lower halves of the stitched plots. An additional, horizontal gradient appears in the $(\frac{1}{2}, \frac{1}{2}, \frac{1}{2})$ peak that coincides with an increase in integrated intensity, indicating added depth-integrated or terraced subdomains. In contrast, the $(\frac{1}{2}, \frac{1}{2}, \frac{1}{4})$ peak does not exhibit a comparable horizontal subdomain structure.
	
	The FWHM plots of each peak further highlight the domain structure, showing an expected broadening near the domain boundaries. Both the FWHM of the $(\frac{1}{2}, \frac{1}{2}, \frac{1}{2})$ and $(\frac{1}{2}, \frac{1}{2}, \frac{1}{4})$ peaks show a slight broadening at the vertical domain boundary identified in the mosaic maps. Furthermore, elliptical domain boundaries appear in the FWHM plots of the $(\frac{1}{2}, \frac{1}{2}, \frac{1}{2})$ peak in the region where a horizontal gradient in the mosaic offset was identified. This indicates the presence of two to three additional domains contained within the lower half of the scan. The FWHM map for the $(\frac{1}{2}, \frac{1}{2}, \frac{1}{4})$ peak remains relatively uniform in this same region (aside from the slight broadening at the vertical domain boundary).
	
	Turning now to the integrated intensity maps of Fig. \ref{fig:CDW}, a clear anticorrelation between the $(\frac{1}{2}, \frac{1}{2}, \frac{1}{2})$ and $(\frac{1}{2}, \frac{1}{2}, \frac{1}{4})$ maps appears. In regions where the $(\frac{1}{2}, \frac{1}{2}, \frac{1}{2})$ peak is maximal, the $(\frac{1}{2}, \frac{1}{2}, \frac{1}{4})$ peak is reduced and visa versa. This is additional evidence that the $(\frac{1}{2}, \frac{1}{2}, \frac{1}{4})$ peak directly competes with the $(\frac{1}{2}, \frac{1}{2}, \frac{1}{2})$ peak regions for domain formation. We stress here, however, that throughout the majority of these maps both peaks appear with varying intensities. This implies a mixed signal via depth averaging of spatially separated domains through the crystal volume. 
	
	
	To further quantify the length scale of the CDW competition, a segmentation plot of the scanned area is presented in Fig. \ref{fig:CDW}. Segmentation was carried out by evaluating the $\theta$-integrated intensity from each CDW signal for a given pixel. The signal that produced a higher integrated intensity is assumed to come from the dominant volume and is assigned a peak label. Regions colored with cyan represent regions where the integrated intensity of the $(\frac{1}{2}, \frac{1}{2}, \frac{1}{4})$ peak exceed the integrated intensity of the $(\frac{1}{2}, \frac{1}{2}, \frac{1}{2})$ peak by 30\%, while orange regions correspond to the inverse condition. Black pixels are designated in regions where integrated intensities are within 30\% of one another. The relative scarcity of black pixels observed in the segmentation plot illustrates that, even when summed along the projection vector, regions prefer to be dominated by one CDW peak or the other. This segregation agrees with coarser resolution scans near the V K-edge in Xiao et al. \cite{xiao2023coexistence}. These results further support the competition and existence of two distinct CDW phases that form separately across crystal domains. 
	
	The mesoscale $(\mu m)$ DFXM results can be compared to macroscale $(mm)$ optical images of the sample to ascertain possible sources for the spatial variation in the CDW state's character. Fig. \ref{fig:ScanVol} displays the scanned area relative to the real-space sample along the $ab$-plane. A relatively continuous metallic surface can be observed over much of the sample area; however the top of the scanned area corresponds to a deformed region that extends over a large void. The deformation manifests as a tab of material (10-50 $\mu m$ thick) still connected to the top most layers of the nominal sample volume.
	
	
	Surveying across this region, where the thickness of the crystal becomes narrower, suggests that changes in the CDW domain structure correspond to defect boundaries in the crystal. The segmented region with a dominant $(\frac{1}{2}, \frac{1}{2}, \frac{1}{4})$ peak appears across the boundary with the upper tab of crystal and seemingly also begins to appear near a lower boundary line at the bottom of the scan. To further elucidate how these surface features correlate with the depth dependence of the CDW states using DFXM, requires overcoming the loss of depth sensitivity along the projection vector of the Bragg peak, which is an active area of research. Despite this region being thinner than the neighboring domain, the crystallinity of the CDW order as assessed via the FWHM of the CDW peaks is identical on either side of the boundary. This suggests a picture of spontaneous nucleation of nearly degenerate CDW states about a local defect rather than a deformation-driven change in the local energy landscape. Regardless of the impetus for the spatial transition between $2\times 2\times 2$ to $2\times 2\times 4$ CDW states, the DXFM data demonstrate a clear competition between the two states in spatially segregated regions at the 100 $\mu m$ length scale.

	\section{Summary}
	The metastability and phase separation of CDW states in CsV$_3$Sb$_5$ were investigated across multiple length scales using X-ray diffraction. 
	Cluster analysis of sample-averaged scattering data shows that multiple CDW states form at offset temperatures and that the onset temperature of CDW order and the mixing between CDW states can be biased via thermal quenching. Data also demonstrate that subtle damage from mechanical cutting quenches the $2\times2\times4$ CDW state, and the metastability associated with thermal hysteresis and competing CDW orders similarly vanishes. DXFM data directly image the spatial separation between CDW domains with dominant $2\times2\times4$ CDW and $2\times2\times2$ CDW characters. These domains seemingly switch along defect boundaries apparent at the crystal surface, suggesting a complex interplay between the nucleation of CDW order and texture (\textit{e.g.} defects, twin domain boundaries) within the crystal lattice. These data are consistent with a number of nearly degenerate CDW states with wave vectors along the M-L line \cite{wu2022charge} that, across a first order phase boundary, are pinned locally by defects in the crystal. Our results stress the importance of understanding the microstructure of CsV$_3$Sb$_5$ in order to interpret the anomalies present in its electronic properties, such as rotational symmetry breaking.

	\section*{Data Availability}
	The raw data that support the findings of this study and the python scripts used for data analysis are openly available in Zenodo at \url{https://doi.org/10.5281/zenodo.12210742}.
	
	\section*{Acknowledgments}
	
	This research was supported by the UC Santa Barbara NSF Quantum Foundry funded via the Q-AMASE-i program under award DMR-1906325. We acknowledge the use of shared facilities of the NSF MRSEC at UC Santa Barbara [DMR 1720256] and the Center for Scientific Computing, supported by the California Nano Systems Institute, the NSF MRSEC (DMR 1720256) and NSF CNS 1725797. This research used resources of the Advanced Photon Source, a U.S. Department of Energy (DOE) Office of Science User Facility at Argonne National Laboratory and is based on research supported by the U.S. DOE Office of Science-Basic Energy Sciences, under Contract No. DE-AC02-06CH11357. Research conducted at the Center for High-Energy X-ray Science (CHEXS) is supported by the NSF (BIO, ENG and MPS Directorates) under award DMR-1829070. The authors would also like to thank the Karlsruhe Nano Micro Facility (KNMF) for the fabrication of the polymer X-ray optics.
	
	K.M. was supported by Eric and Wendy Schmidt AI in Science Postdoctoral Fellowship: a Schmidt Futures program. The X-TEC analysis was carried out on a high-powered computing cluster funded in part by the New Frontier Grant from the College of Arts and Sciences at Cornell (to E.-A.K.) and by the Gordon and Betty Moore Foundation’s EPiQS Initiative via Grant No. GBMF10436 (to E.-A.K.).
	
	F.B.C. was supported by the U.S. Department of Energy, Office of Basic Energy Sciences, Division of Materials Science and Engineering project "Quantum Fluctuations in Narrow-Band Systems" and the Coordenação de Aperfeiçoamento de Pessoal de Nível Superior - Brasil (CAPES) - Finance Code 001.

	\bibliography{CVS_Plumb_Refs}

\begin{thebibliography}{39}%
\makeatletter
\providecommand \@ifxundefined [1]{%
 \@ifx{#1\undefined}
}%
\providecommand \@ifnum [1]{%
 \ifnum #1\expandafter \@firstoftwo
 \else \expandafter \@secondoftwo
 \fi
}%
\providecommand \@ifx [1]{%
 \ifx #1\expandafter \@firstoftwo
 \else \expandafter \@secondoftwo
 \fi
}%
\providecommand \natexlab [1]{#1}%
\providecommand \enquote  [1]{``#1''}%
\providecommand \bibnamefont  [1]{#1}%
\providecommand \bibfnamefont [1]{#1}%
\providecommand \citenamefont [1]{#1}%
\providecommand \href@noop [0]{\@secondoftwo}%
\providecommand \href [0]{\begingroup \@sanitize@url \@href}%
\providecommand \@href[1]{\@@startlink{#1}\@@href}%
\providecommand \@@href[1]{\endgroup#1\@@endlink}%
\providecommand \@sanitize@url [0]{\catcode `\\12\catcode `\$12\catcode
  `\&12\catcode `\#12\catcode `\^12\catcode `\_12\catcode `\%12\relax}%
\providecommand \@@startlink[1]{}%
\providecommand \@@endlink[0]{}%
\providecommand \url  [0]{\begingroup\@sanitize@url \@url }%
\providecommand \@url [1]{\endgroup\@href {#1}{\urlprefix }}%
\providecommand \urlprefix  [0]{URL }%
\providecommand \Eprint [0]{\href }%
\providecommand \doibase [0]{https://doi.org/}%
\providecommand \selectlanguage [0]{\@gobble}%
\providecommand \bibinfo  [0]{\@secondoftwo}%
\providecommand \bibfield  [0]{\@secondoftwo}%
\providecommand \translation [1]{[#1]}%
\providecommand \BibitemOpen [0]{}%
\providecommand \bibitemStop [0]{}%
\providecommand \bibitemNoStop [0]{.\EOS\space}%
\providecommand \EOS [0]{\spacefactor3000\relax}%
\providecommand \BibitemShut  [1]{\csname bibitem#1\endcsname}%
\let\auto@bib@innerbib\@empty
\bibitem [{\citenamefont {Ortiz}\ \emph {et~al.}(2019)\citenamefont {Ortiz},
  \citenamefont {Gomes}, \citenamefont {Morey}, \citenamefont {Winiarski},
  \citenamefont {Bordelon}, \citenamefont {Mangum}, \citenamefont {Oswald},
  \citenamefont {Rodriguez-Rivera}, \citenamefont {Neilson}, \citenamefont
  {Wilson}, \citenamefont {Ertekin}, \citenamefont {McQueen},\ and\
  \citenamefont {Toberer}}]{ortiz2019new}%
  \BibitemOpen
  \bibfield  {author} {\bibinfo {author} {\bibfnamefont {B.~R.}\ \bibnamefont
  {Ortiz}}, \bibinfo {author} {\bibfnamefont {L.~C.}\ \bibnamefont {Gomes}},
  \bibinfo {author} {\bibfnamefont {J.~R.}\ \bibnamefont {Morey}}, \bibinfo
  {author} {\bibfnamefont {M.}~\bibnamefont {Winiarski}}, \bibinfo {author}
  {\bibfnamefont {M.}~\bibnamefont {Bordelon}}, \bibinfo {author}
  {\bibfnamefont {J.~S.}\ \bibnamefont {Mangum}}, \bibinfo {author}
  {\bibfnamefont {I.~W.~H.}\ \bibnamefont {Oswald}}, \bibinfo {author}
  {\bibfnamefont {J.~A.}\ \bibnamefont {Rodriguez-Rivera}}, \bibinfo {author}
  {\bibfnamefont {J.~R.}\ \bibnamefont {Neilson}}, \bibinfo {author}
  {\bibfnamefont {S.~D.}\ \bibnamefont {Wilson}}, \bibinfo {author}
  {\bibfnamefont {E.}~\bibnamefont {Ertekin}}, \bibinfo {author} {\bibfnamefont
  {T.~M.}\ \bibnamefont {McQueen}},\ and\ \bibinfo {author} {\bibfnamefont
  {E.~S.}\ \bibnamefont {Toberer}},\ }\bibfield  {title} {\bibinfo {title} {New
  kagome prototype materials: discovery of \( \mathrm{KV_3Sb_5} \), \(
  \mathrm{RbV_3Sb_5} \), and \( \mathrm{CsV_3Sb_5} \)},\ }\href
  {https://doi.org/10.1103/PhysRevMaterials.3.094407} {\bibfield  {journal}
  {\bibinfo  {journal} {Physical Review Materials}\ }\textbf {\bibinfo {volume}
  {3}},\ \bibinfo {pages} {094407} (\bibinfo {year} {2019})}\BibitemShut
  {NoStop}%
\bibitem [{\citenamefont {Yang}\ \emph {et~al.}(2020)\citenamefont {Yang},
  \citenamefont {Wang}, \citenamefont {Ortiz}, \citenamefont {Liu},
  \citenamefont {Gayles}, \citenamefont {Derunova}, \citenamefont
  {Gonzalez-Hernandez}, \citenamefont {Šmejkal}, \citenamefont {Chen},
  \citenamefont {Parkin}, \citenamefont {Wilson}, \citenamefont {Toberer},
  \citenamefont {McQueen},\ and\ \citenamefont {Ali}}]{yang2020giant}%
  \BibitemOpen
  \bibfield  {author} {\bibinfo {author} {\bibfnamefont {S.-Y.}\ \bibnamefont
  {Yang}}, \bibinfo {author} {\bibfnamefont {Y.}~\bibnamefont {Wang}}, \bibinfo
  {author} {\bibfnamefont {B.~R.}\ \bibnamefont {Ortiz}}, \bibinfo {author}
  {\bibfnamefont {D.}~\bibnamefont {Liu}}, \bibinfo {author} {\bibfnamefont
  {J.}~\bibnamefont {Gayles}}, \bibinfo {author} {\bibfnamefont
  {E.}~\bibnamefont {Derunova}}, \bibinfo {author} {\bibfnamefont
  {R.}~\bibnamefont {Gonzalez-Hernandez}}, \bibinfo {author} {\bibfnamefont
  {L.}~\bibnamefont {Šmejkal}}, \bibinfo {author} {\bibfnamefont
  {Y.}~\bibnamefont {Chen}}, \bibinfo {author} {\bibfnamefont {S.~S.~P.}\
  \bibnamefont {Parkin}}, \bibinfo {author} {\bibfnamefont {S.~D.}\
  \bibnamefont {Wilson}}, \bibinfo {author} {\bibfnamefont {E.~S.}\
  \bibnamefont {Toberer}}, \bibinfo {author} {\bibfnamefont {T.}~\bibnamefont
  {McQueen}},\ and\ \bibinfo {author} {\bibfnamefont {M.~N.}\ \bibnamefont
  {Ali}},\ }\bibfield  {title} {\bibinfo {title} {Giant, unconventional
  anomalous hall effect in the metallic frustrated magnet candidate, \(
  \mathrm{KV_3Sb_5} \)},\ }\href {https://doi.org/10.1126/sciadv.abb6003}
  {\bibfield  {journal} {\bibinfo  {journal} {Science advances}\ }\textbf
  {\bibinfo {volume} {6}},\ \bibinfo {pages} {eabb6003} (\bibinfo {year}
  {2020})}\BibitemShut {NoStop}%
\bibitem [{\citenamefont {Yu}\ \emph {et~al.}(2021)\citenamefont {Yu},
  \citenamefont {Wu}, \citenamefont {Wang}, \citenamefont {Lei}, \citenamefont
  {Zhuo}, \citenamefont {Ying},\ and\ \citenamefont
  {Chen}}]{yu2021concurrence}%
  \BibitemOpen
  \bibfield  {author} {\bibinfo {author} {\bibfnamefont {F.~H.}\ \bibnamefont
  {Yu}}, \bibinfo {author} {\bibfnamefont {T.}~\bibnamefont {Wu}}, \bibinfo
  {author} {\bibfnamefont {Z.~Y.}\ \bibnamefont {Wang}}, \bibinfo {author}
  {\bibfnamefont {B.}~\bibnamefont {Lei}}, \bibinfo {author} {\bibfnamefont
  {W.~Z.}\ \bibnamefont {Zhuo}}, \bibinfo {author} {\bibfnamefont {J.~J.}\
  \bibnamefont {Ying}},\ and\ \bibinfo {author} {\bibfnamefont {X.~H.}\
  \bibnamefont {Chen}},\ }\bibfield  {title} {\bibinfo {title} {Concurrence of
  anomalous hall effect and charge density wave in a superconducting
  topological kagome metal},\ }\href
  {https://doi.org/10.1103/PhysRevB.104.L041103} {\bibfield  {journal}
  {\bibinfo  {journal} {Physical Review B}\ }\textbf {\bibinfo {volume}
  {104}},\ \bibinfo {pages} {L041103} (\bibinfo {year} {2021})}\BibitemShut
  {NoStop}%
\bibitem [{\citenamefont {Zhao}\ \emph {et~al.}(2021)\citenamefont {Zhao},
  \citenamefont {Li}, \citenamefont {Ortiz}, \citenamefont {Teicher},
  \citenamefont {Park}, \citenamefont {Ye}, \citenamefont {Wang}, \citenamefont
  {Balents}, \citenamefont {Wilson},\ and\ \citenamefont
  {Zeljkovic}}]{zhao2021cascade}%
  \BibitemOpen
  \bibfield  {author} {\bibinfo {author} {\bibfnamefont {H.}~\bibnamefont
  {Zhao}}, \bibinfo {author} {\bibfnamefont {H.}~\bibnamefont {Li}}, \bibinfo
  {author} {\bibfnamefont {B.~R.}\ \bibnamefont {Ortiz}}, \bibinfo {author}
  {\bibfnamefont {S.~M.}\ \bibnamefont {Teicher}}, \bibinfo {author}
  {\bibfnamefont {T.}~\bibnamefont {Park}}, \bibinfo {author} {\bibfnamefont
  {M.}~\bibnamefont {Ye}}, \bibinfo {author} {\bibfnamefont {Z.}~\bibnamefont
  {Wang}}, \bibinfo {author} {\bibfnamefont {L.}~\bibnamefont {Balents}},
  \bibinfo {author} {\bibfnamefont {S.~D.}\ \bibnamefont {Wilson}},\ and\
  \bibinfo {author} {\bibfnamefont {I.}~\bibnamefont {Zeljkovic}},\ }\bibfield
  {title} {\bibinfo {title} {Cascade of correlated electron states in the
  kagome superconductor \( \mathrm{CsV_3Sb_5} \)},\ }\href
  {https://doi.org/10.1038/s41586-021-03946-w} {\bibfield  {journal} {\bibinfo
  {journal} {Nature}\ }\textbf {\bibinfo {volume} {599}},\ \bibinfo {pages}
  {216} (\bibinfo {year} {2021})}\BibitemShut {NoStop}%
\bibitem [{\citenamefont {Li}\ \emph {et~al.}(2023)\citenamefont {Li},
  \citenamefont {Zhao}, \citenamefont {Ortiz}, \citenamefont {Oey},
  \citenamefont {Wang}, \citenamefont {Wilson},\ and\ \citenamefont
  {Zeljkovic}}]{li2023unidirectional}%
  \BibitemOpen
  \bibfield  {author} {\bibinfo {author} {\bibfnamefont {H.}~\bibnamefont
  {Li}}, \bibinfo {author} {\bibfnamefont {H.}~\bibnamefont {Zhao}}, \bibinfo
  {author} {\bibfnamefont {B.~R.}\ \bibnamefont {Ortiz}}, \bibinfo {author}
  {\bibfnamefont {Y.}~\bibnamefont {Oey}}, \bibinfo {author} {\bibfnamefont
  {Z.}~\bibnamefont {Wang}}, \bibinfo {author} {\bibfnamefont {S.~D.}\
  \bibnamefont {Wilson}},\ and\ \bibinfo {author} {\bibfnamefont
  {I.}~\bibnamefont {Zeljkovic}},\ }\bibfield  {title} {\bibinfo {title}
  {Unidirectional coherent quasiparticles in the high-temperature rotational
  symmetry broken phase of \( \mathrm{AV_3Sb_5} \) kagome superconductors},\
  }\href {https://doi.org/10.1038/s41567-022-01932-1} {\bibfield  {journal}
  {\bibinfo  {journal} {Nature Physics}\ }\textbf {\bibinfo {volume} {19}},\
  \bibinfo {pages} {637} (\bibinfo {year} {2023})}\BibitemShut {NoStop}%
\bibitem [{\citenamefont {Luo}\ \emph {et~al.}(2022)\citenamefont {Luo},
  \citenamefont {Zhao}, \citenamefont {Zhou}, \citenamefont {Yang},
  \citenamefont {Fang}, \citenamefont {Yang}, \citenamefont {Gao},
  \citenamefont {Zhou},\ and\ \citenamefont {Zheng}}]{luo2022possible}%
  \BibitemOpen
  \bibfield  {author} {\bibinfo {author} {\bibfnamefont {J.}~\bibnamefont
  {Luo}}, \bibinfo {author} {\bibfnamefont {Z.}~\bibnamefont {Zhao}}, \bibinfo
  {author} {\bibfnamefont {Y.}~\bibnamefont {Zhou}}, \bibinfo {author}
  {\bibfnamefont {J.}~\bibnamefont {Yang}}, \bibinfo {author} {\bibfnamefont
  {A.}~\bibnamefont {Fang}}, \bibinfo {author} {\bibfnamefont {H.}~\bibnamefont
  {Yang}}, \bibinfo {author} {\bibfnamefont {H.}~\bibnamefont {Gao}}, \bibinfo
  {author} {\bibfnamefont {R.}~\bibnamefont {Zhou}},\ and\ \bibinfo {author}
  {\bibfnamefont {G.-q.}\ \bibnamefont {Zheng}},\ }\bibfield  {title} {\bibinfo
  {title} {Possible star-of-david pattern charge density wave with additional
  modulation in the kagome superconductor \( \mathrm{CsV_3Sb_5} \)},\ }\href
  {https://doi.org/10.1038/s41535-022-00437-7} {\bibfield  {journal} {\bibinfo
  {journal} {npj Quantum Materials}\ }\textbf {\bibinfo {volume} {7}},\
  \bibinfo {pages} {30} (\bibinfo {year} {2022})}\BibitemShut {NoStop}%
\bibitem [{\citenamefont {Wilson}\ and\ \citenamefont
  {Ortiz}(2024)}]{Wilson2024}%
  \BibitemOpen
  \bibfield  {author} {\bibinfo {author} {\bibfnamefont {S.~D.}\ \bibnamefont
  {Wilson}}\ and\ \bibinfo {author} {\bibfnamefont {B.~R.}\ \bibnamefont
  {Ortiz}},\ }\bibfield  {title} {\bibinfo {title} {\( \mathrm{AV_3Sb_5} \)
  kagome superconductors},\ }\href {https://doi.org/10.1038/s41578-024-00677-y}
  {\bibfield  {journal} {\bibinfo  {journal} {Nature Reviews Materials}\
  }\textbf {\bibinfo {volume} {9}},\ \bibinfo {pages} {420} (\bibinfo {year}
  {2024})}\BibitemShut {NoStop}%
\bibitem [{\citenamefont {Jiang}\ \emph {et~al.}(2021)\citenamefont {Jiang},
  \citenamefont {Yin}, \citenamefont {Denner}, \citenamefont {Shumiya},
  \citenamefont {Ortiz}, \citenamefont {Xu}, \citenamefont {Guguchia},
  \citenamefont {He}, \citenamefont {Hossain}, \citenamefont {Liu},
  \citenamefont {Ruff}, \citenamefont {Kautzsch}, \citenamefont {Zhang},
  \citenamefont {Chang}, \citenamefont {Belopolski}, \citenamefont {Zhang},
  \citenamefont {Cochran}, \citenamefont {Multer}, \citenamefont {Litskevich},
  \citenamefont {Cheng}, \citenamefont {Yang}, \citenamefont {Wang},
  \citenamefont {Thomale}, \citenamefont {Neupert}, \citenamefont {Wilson},\
  and\ \citenamefont {Hasan}}]{jiang2021unconventional}%
  \BibitemOpen
  \bibfield  {author} {\bibinfo {author} {\bibfnamefont {Y.-X.}\ \bibnamefont
  {Jiang}}, \bibinfo {author} {\bibfnamefont {J.-X.}\ \bibnamefont {Yin}},
  \bibinfo {author} {\bibfnamefont {M.~M.}\ \bibnamefont {Denner}}, \bibinfo
  {author} {\bibfnamefont {N.}~\bibnamefont {Shumiya}}, \bibinfo {author}
  {\bibfnamefont {B.~R.}\ \bibnamefont {Ortiz}}, \bibinfo {author}
  {\bibfnamefont {G.}~\bibnamefont {Xu}}, \bibinfo {author} {\bibfnamefont
  {Z.}~\bibnamefont {Guguchia}}, \bibinfo {author} {\bibfnamefont
  {J.}~\bibnamefont {He}}, \bibinfo {author} {\bibfnamefont {M.~S.}\
  \bibnamefont {Hossain}}, \bibinfo {author} {\bibfnamefont {X.}~\bibnamefont
  {Liu}}, \bibinfo {author} {\bibfnamefont {J.}~\bibnamefont {Ruff}}, \bibinfo
  {author} {\bibfnamefont {L.}~\bibnamefont {Kautzsch}}, \bibinfo {author}
  {\bibfnamefont {S.~S.}\ \bibnamefont {Zhang}}, \bibinfo {author}
  {\bibfnamefont {G.}~\bibnamefont {Chang}}, \bibinfo {author} {\bibfnamefont
  {I.}~\bibnamefont {Belopolski}}, \bibinfo {author} {\bibfnamefont
  {Q.}~\bibnamefont {Zhang}}, \bibinfo {author} {\bibfnamefont {T.~A.}\
  \bibnamefont {Cochran}}, \bibinfo {author} {\bibfnamefont {D.}~\bibnamefont
  {Multer}}, \bibinfo {author} {\bibfnamefont {M.}~\bibnamefont {Litskevich}},
  \bibinfo {author} {\bibfnamefont {Z.-J.}\ \bibnamefont {Cheng}}, \bibinfo
  {author} {\bibfnamefont {X.~P.}\ \bibnamefont {Yang}}, \bibinfo {author}
  {\bibfnamefont {Z.}~\bibnamefont {Wang}}, \bibinfo {author} {\bibfnamefont
  {R.}~\bibnamefont {Thomale}}, \bibinfo {author} {\bibfnamefont
  {T.}~\bibnamefont {Neupert}}, \bibinfo {author} {\bibfnamefont {S.~D.}\
  \bibnamefont {Wilson}},\ and\ \bibinfo {author} {\bibfnamefont {M.~Z.}\
  \bibnamefont {Hasan}},\ }\bibfield  {title} {\bibinfo {title} {Unconventional
  chiral charge order in kagome superconductor \( \mathrm{KV_3Sb_5} \)},\
  }\href {https://doi.org/10.1038/s41563-021-01034-y} {\bibfield  {journal}
  {\bibinfo  {journal} {Nature materials}\ }\textbf {\bibinfo {volume} {20}},\
  \bibinfo {pages} {1353} (\bibinfo {year} {2021})}\BibitemShut {NoStop}%
\bibitem [{\citenamefont {Mielke}\ \emph {et~al.}(2022)\citenamefont {Mielke},
  \citenamefont {Das}, \citenamefont {Yin}, \citenamefont {Liu}, \citenamefont
  {Gupta}, \citenamefont {Jiang}, \citenamefont {Medarde}, \citenamefont {Wu},
  \citenamefont {Lei}, \citenamefont {Chang}, \citenamefont {Dai},
  \citenamefont {Si}, \citenamefont {Miao}, \citenamefont {Thomale},
  \citenamefont {Neupert}, \citenamefont {Shi}, \citenamefont {Khasanov},
  \citenamefont {Hasan}, \citenamefont {Luetkens},\ and\ \citenamefont
  {Guguchia}}]{mielke2022time}%
  \BibitemOpen
  \bibfield  {author} {\bibinfo {author} {\bibfnamefont {C.}~\bibnamefont
  {Mielke}}, \bibinfo {author} {\bibfnamefont {D.}~\bibnamefont {Das}},
  \bibinfo {author} {\bibfnamefont {J.-X.}\ \bibnamefont {Yin}}, \bibinfo
  {author} {\bibfnamefont {H.}~\bibnamefont {Liu}}, \bibinfo {author}
  {\bibfnamefont {R.}~\bibnamefont {Gupta}}, \bibinfo {author} {\bibfnamefont
  {Y.-X.}\ \bibnamefont {Jiang}}, \bibinfo {author} {\bibfnamefont
  {M.}~\bibnamefont {Medarde}}, \bibinfo {author} {\bibfnamefont
  {X.}~\bibnamefont {Wu}}, \bibinfo {author} {\bibfnamefont {H.~C.}\
  \bibnamefont {Lei}}, \bibinfo {author} {\bibfnamefont {J.}~\bibnamefont
  {Chang}}, \bibinfo {author} {\bibfnamefont {P.}~\bibnamefont {Dai}}, \bibinfo
  {author} {\bibfnamefont {Q.}~\bibnamefont {Si}}, \bibinfo {author}
  {\bibfnamefont {H.}~\bibnamefont {Miao}}, \bibinfo {author} {\bibfnamefont
  {R.}~\bibnamefont {Thomale}}, \bibinfo {author} {\bibfnamefont
  {T.}~\bibnamefont {Neupert}}, \bibinfo {author} {\bibfnamefont
  {Y.}~\bibnamefont {Shi}}, \bibinfo {author} {\bibfnamefont {R.}~\bibnamefont
  {Khasanov}}, \bibinfo {author} {\bibfnamefont {M.~Z.}\ \bibnamefont {Hasan}},
  \bibinfo {author} {\bibfnamefont {H.}~\bibnamefont {Luetkens}},\ and\
  \bibinfo {author} {\bibfnamefont {Z.}~\bibnamefont {Guguchia}},\ }\bibfield
  {title} {\bibinfo {title} {Time-reversal symmetry-breaking charge order in a
  kagome superconductor},\ }\href {https://doi.org/10.1038/s41586-021-04327-z}
  {\bibfield  {journal} {\bibinfo  {journal} {Nature}\ }\textbf {\bibinfo
  {volume} {602}},\ \bibinfo {pages} {245} (\bibinfo {year}
  {2022})}\BibitemShut {NoStop}%
\bibitem [{\citenamefont {Xing}\ \emph {et~al.}(2024)\citenamefont {Xing},
  \citenamefont {Bae}, \citenamefont {Ritz}, \citenamefont {Yang},
  \citenamefont {Birol}, \citenamefont {Capa~Salinas}, \citenamefont {Ortiz},
  \citenamefont {Wilson}, \citenamefont {Wang}, \citenamefont {Fernandes},\
  and\ \citenamefont {Madhavan}}]{xing2024optical}%
  \BibitemOpen
  \bibfield  {author} {\bibinfo {author} {\bibfnamefont {Y.}~\bibnamefont
  {Xing}}, \bibinfo {author} {\bibfnamefont {S.}~\bibnamefont {Bae}}, \bibinfo
  {author} {\bibfnamefont {E.}~\bibnamefont {Ritz}}, \bibinfo {author}
  {\bibfnamefont {F.}~\bibnamefont {Yang}}, \bibinfo {author} {\bibfnamefont
  {T.}~\bibnamefont {Birol}}, \bibinfo {author} {\bibfnamefont {A.~N.}\
  \bibnamefont {Capa~Salinas}}, \bibinfo {author} {\bibfnamefont {B.~R.}\
  \bibnamefont {Ortiz}}, \bibinfo {author} {\bibfnamefont {S.~D.}\ \bibnamefont
  {Wilson}}, \bibinfo {author} {\bibfnamefont {Z.}~\bibnamefont {Wang}},
  \bibinfo {author} {\bibfnamefont {R.~M.}\ \bibnamefont {Fernandes}},\ and\
  \bibinfo {author} {\bibfnamefont {V.}~\bibnamefont {Madhavan}},\ }\bibfield
  {title} {\bibinfo {title} {Optical manipulation of the charge-density-wave
  state in \( \mathrm{RbV_3Sb_5} \)},\ }\href
  {https://doi.org/10.1038/s41586-024-07519-5} {\bibfield  {journal} {\bibinfo
  {journal} {Nature}\ ,\ \bibinfo {pages} {1}} (\bibinfo {year}
  {2024})}\BibitemShut {NoStop}%
\bibitem [{\citenamefont {Le}\ \emph {et~al.}(2024)\citenamefont {Le},
  \citenamefont {Pan}, \citenamefont {Xu}, \citenamefont {Liu}, \citenamefont
  {Wang}, \citenamefont {Lou}, \citenamefont {Yang}, \citenamefont {Wang},
  \citenamefont {Yao}, \citenamefont {Wu},\ and\ \citenamefont
  {Lin}}]{le2024superconducting}%
  \BibitemOpen
  \bibfield  {author} {\bibinfo {author} {\bibfnamefont {T.}~\bibnamefont
  {Le}}, \bibinfo {author} {\bibfnamefont {Z.}~\bibnamefont {Pan}}, \bibinfo
  {author} {\bibfnamefont {Z.}~\bibnamefont {Xu}}, \bibinfo {author}
  {\bibfnamefont {J.}~\bibnamefont {Liu}}, \bibinfo {author} {\bibfnamefont
  {J.}~\bibnamefont {Wang}}, \bibinfo {author} {\bibfnamefont {Z.}~\bibnamefont
  {Lou}}, \bibinfo {author} {\bibfnamefont {X.}~\bibnamefont {Yang}}, \bibinfo
  {author} {\bibfnamefont {Z.}~\bibnamefont {Wang}}, \bibinfo {author}
  {\bibfnamefont {Y.}~\bibnamefont {Yao}}, \bibinfo {author} {\bibfnamefont
  {C.}~\bibnamefont {Wu}},\ and\ \bibinfo {author} {\bibfnamefont
  {X.}~\bibnamefont {Lin}},\ }\bibfield  {title} {\bibinfo {title}
  {Superconducting diode effect and interference patterns in kagome \(
  \mathrm{CsV_3Sb_5} \)},\ }\href {https://doi.org/10.1038/s41586-024-07431-y}
  {\bibfield  {journal} {\bibinfo  {journal} {Nature}\ ,\ \bibinfo {pages} {1}}
  (\bibinfo {year} {2024})}\BibitemShut {NoStop}%
\bibitem [{\citenamefont {Xu}\ \emph {et~al.}(2022)\citenamefont {Xu},
  \citenamefont {Ni}, \citenamefont {Liu}, \citenamefont {Ortiz}, \citenamefont
  {Deng}, \citenamefont {Wilson}, \citenamefont {Yan}, \citenamefont
  {Balents},\ and\ \citenamefont {Wu}}]{xu2022three}%
  \BibitemOpen
  \bibfield  {author} {\bibinfo {author} {\bibfnamefont {Y.}~\bibnamefont
  {Xu}}, \bibinfo {author} {\bibfnamefont {Z.}~\bibnamefont {Ni}}, \bibinfo
  {author} {\bibfnamefont {Y.}~\bibnamefont {Liu}}, \bibinfo {author}
  {\bibfnamefont {B.~R.}\ \bibnamefont {Ortiz}}, \bibinfo {author}
  {\bibfnamefont {Q.}~\bibnamefont {Deng}}, \bibinfo {author} {\bibfnamefont
  {S.~D.}\ \bibnamefont {Wilson}}, \bibinfo {author} {\bibfnamefont
  {B.}~\bibnamefont {Yan}}, \bibinfo {author} {\bibfnamefont {L.}~\bibnamefont
  {Balents}},\ and\ \bibinfo {author} {\bibfnamefont {L.}~\bibnamefont {Wu}},\
  }\bibfield  {title} {\bibinfo {title} {Three-state nematicity and
  magneto-optical kerr effect in the charge density waves in kagome
  superconductors},\ }\href {https://doi.org/10.1038/s41567-022-01805-7}
  {\bibfield  {journal} {\bibinfo  {journal} {Nature physics}\ }\textbf
  {\bibinfo {volume} {18}},\ \bibinfo {pages} {1470} (\bibinfo {year}
  {2022})}\BibitemShut {NoStop}%
\bibitem [{\citenamefont {Ortiz}\ \emph {et~al.}(2020)\citenamefont {Ortiz},
  \citenamefont {Teicher}, \citenamefont {Hu}, \citenamefont {Zuo},
  \citenamefont {Sarte}, \citenamefont {Schueller}, \citenamefont {Abeykoon},
  \citenamefont {Krogstad}, \citenamefont {Rosenkranz}, \citenamefont {Osborn},
  \citenamefont {Seshadri}, \citenamefont {Balents}, \citenamefont {He},\ and\
  \citenamefont {Wilson}}]{ortiz2020cs}%
  \BibitemOpen
  \bibfield  {author} {\bibinfo {author} {\bibfnamefont {B.~R.}\ \bibnamefont
  {Ortiz}}, \bibinfo {author} {\bibfnamefont {S.~M.~L.}\ \bibnamefont
  {Teicher}}, \bibinfo {author} {\bibfnamefont {Y.}~\bibnamefont {Hu}},
  \bibinfo {author} {\bibfnamefont {J.~L.}\ \bibnamefont {Zuo}}, \bibinfo
  {author} {\bibfnamefont {P.~M.}\ \bibnamefont {Sarte}}, \bibinfo {author}
  {\bibfnamefont {E.~C.}\ \bibnamefont {Schueller}}, \bibinfo {author}
  {\bibfnamefont {A.~M.~M.}\ \bibnamefont {Abeykoon}}, \bibinfo {author}
  {\bibfnamefont {M.~J.}\ \bibnamefont {Krogstad}}, \bibinfo {author}
  {\bibfnamefont {S.}~\bibnamefont {Rosenkranz}}, \bibinfo {author}
  {\bibfnamefont {R.}~\bibnamefont {Osborn}}, \bibinfo {author} {\bibfnamefont
  {R.}~\bibnamefont {Seshadri}}, \bibinfo {author} {\bibfnamefont
  {L.}~\bibnamefont {Balents}}, \bibinfo {author} {\bibfnamefont
  {J.}~\bibnamefont {He}},\ and\ \bibinfo {author} {\bibfnamefont {S.~D.}\
  \bibnamefont {Wilson}},\ }\bibfield  {title} {\bibinfo {title}
  {\(\mathrm{CsV_3Sb_5} \): A \( \mathrm{Z_{2}} \) topological kagome metal
  with a superconducting ground state},\ }\href
  {https://doi.org/10.1103/PhysRevLett.125.247002} {\bibfield  {journal}
  {\bibinfo  {journal} {Physical Review Letters}\ }\textbf {\bibinfo {volume}
  {125}},\ \bibinfo {pages} {247002} (\bibinfo {year} {2020})}\BibitemShut
  {NoStop}%
\bibitem [{\citenamefont {Ortiz}\ \emph
  {et~al.}(2021{\natexlab{a}})\citenamefont {Ortiz}, \citenamefont {Sarte},
  \citenamefont {Kenney}, \citenamefont {Graf}, \citenamefont {Teicher},
  \citenamefont {Seshadri},\ and\ \citenamefont
  {Wilson}}]{ortiz2021superconductivity}%
  \BibitemOpen
  \bibfield  {author} {\bibinfo {author} {\bibfnamefont {B.~R.}\ \bibnamefont
  {Ortiz}}, \bibinfo {author} {\bibfnamefont {P.~M.}\ \bibnamefont {Sarte}},
  \bibinfo {author} {\bibfnamefont {E.~M.}\ \bibnamefont {Kenney}}, \bibinfo
  {author} {\bibfnamefont {M.~J.}\ \bibnamefont {Graf}}, \bibinfo {author}
  {\bibfnamefont {S.~M.}\ \bibnamefont {Teicher}}, \bibinfo {author}
  {\bibfnamefont {R.}~\bibnamefont {Seshadri}},\ and\ \bibinfo {author}
  {\bibfnamefont {S.~D.}\ \bibnamefont {Wilson}},\ }\bibfield  {title}
  {\bibinfo {title} {Superconductivity in the \( \mathrm{Z_{2}} \) kagome metal
  \( \mathrm{KV_3Sb_5} \)},\ }\href
  {https://doi.org/10.1103/PhysRevMaterials.5.034801} {\bibfield  {journal}
  {\bibinfo  {journal} {Physical Review Materials}\ }\textbf {\bibinfo {volume}
  {5}},\ \bibinfo {pages} {034801} (\bibinfo {year}
  {2021}{\natexlab{a}})}\BibitemShut {NoStop}%
\bibitem [{\citenamefont {Yin}\ \emph {et~al.}(2021)\citenamefont {Yin},
  \citenamefont {Tu}, \citenamefont {Gong}, \citenamefont {Fu}, \citenamefont
  {Yan},\ and\ \citenamefont {Lei}}]{yin2021superconductivity}%
  \BibitemOpen
  \bibfield  {author} {\bibinfo {author} {\bibfnamefont {Q.}~\bibnamefont
  {Yin}}, \bibinfo {author} {\bibfnamefont {Z.}~\bibnamefont {Tu}}, \bibinfo
  {author} {\bibfnamefont {C.}~\bibnamefont {Gong}}, \bibinfo {author}
  {\bibfnamefont {Y.}~\bibnamefont {Fu}}, \bibinfo {author} {\bibfnamefont
  {S.}~\bibnamefont {Yan}},\ and\ \bibinfo {author} {\bibfnamefont
  {H.}~\bibnamefont {Lei}},\ }\bibfield  {title} {\bibinfo {title}
  {Superconductivity and normal-state properties of kagome metal \(
  \mathrm{RbV_3Sb_5} \) single crystals},\ }\href
  {https://doi.org/10.1088/0256-307X/38/3/037403} {\bibfield  {journal}
  {\bibinfo  {journal} {Chinese Physics Letters}\ }\textbf {\bibinfo {volume}
  {38}},\ \bibinfo {pages} {037403} (\bibinfo {year} {2021})}\BibitemShut
  {NoStop}%
\bibitem [{\citenamefont {Frassineti}\ \emph {et~al.}(2023)\citenamefont
  {Frassineti}, \citenamefont {Bonf{\`a}}, \citenamefont {Allodi},
  \citenamefont {Garcia}, \citenamefont {Cong}, \citenamefont {Ortiz},
  \citenamefont {Wilson}, \citenamefont {De~Renzi}, \citenamefont
  {Mitrovi{\'c}},\ and\ \citenamefont {Sanna}}]{frassineti2023microscopic}%
  \BibitemOpen
  \bibfield  {author} {\bibinfo {author} {\bibfnamefont {J.}~\bibnamefont
  {Frassineti}}, \bibinfo {author} {\bibfnamefont {P.}~\bibnamefont
  {Bonf{\`a}}}, \bibinfo {author} {\bibfnamefont {G.}~\bibnamefont {Allodi}},
  \bibinfo {author} {\bibfnamefont {E.}~\bibnamefont {Garcia}}, \bibinfo
  {author} {\bibfnamefont {R.}~\bibnamefont {Cong}}, \bibinfo {author}
  {\bibfnamefont {B.~R.}\ \bibnamefont {Ortiz}}, \bibinfo {author}
  {\bibfnamefont {S.~D.}\ \bibnamefont {Wilson}}, \bibinfo {author}
  {\bibfnamefont {R.}~\bibnamefont {De~Renzi}}, \bibinfo {author}
  {\bibfnamefont {V.~F.}\ \bibnamefont {Mitrovi{\'c}}},\ and\ \bibinfo {author}
  {\bibfnamefont {S.}~\bibnamefont {Sanna}},\ }\bibfield  {title} {\bibinfo
  {title} {Microscopic nature of the charge-density wave in the kagome
  superconductor \( \mathrm{RbV_3Sb_5} \)},\ }\href
  {https://doi.org/10.1103/PhysRevResearch.5.L012017} {\bibfield  {journal}
  {\bibinfo  {journal} {Physical Review Research}\ }\textbf {\bibinfo {volume}
  {5}},\ \bibinfo {pages} {L012017} (\bibinfo {year} {2023})}\BibitemShut
  {NoStop}%
\bibitem [{\citenamefont {Kautzsch}\ \emph
  {et~al.}(2023{\natexlab{a}})\citenamefont {Kautzsch}, \citenamefont {Ortiz},
  \citenamefont {Mallayya}, \citenamefont {Plumb}, \citenamefont {Pokharel},
  \citenamefont {Ruff}, \citenamefont {Islam}, \citenamefont {Kim},
  \citenamefont {Seshadri},\ and\ \citenamefont
  {Wilson}}]{kautzsch2023structural}%
  \BibitemOpen
  \bibfield  {author} {\bibinfo {author} {\bibfnamefont {L.}~\bibnamefont
  {Kautzsch}}, \bibinfo {author} {\bibfnamefont {B.~R.}\ \bibnamefont {Ortiz}},
  \bibinfo {author} {\bibfnamefont {K.}~\bibnamefont {Mallayya}}, \bibinfo
  {author} {\bibfnamefont {J.}~\bibnamefont {Plumb}}, \bibinfo {author}
  {\bibfnamefont {G.}~\bibnamefont {Pokharel}}, \bibinfo {author}
  {\bibfnamefont {J.~P.}\ \bibnamefont {Ruff}}, \bibinfo {author}
  {\bibfnamefont {Z.}~\bibnamefont {Islam}}, \bibinfo {author} {\bibfnamefont
  {E.-A.}\ \bibnamefont {Kim}}, \bibinfo {author} {\bibfnamefont
  {R.}~\bibnamefont {Seshadri}},\ and\ \bibinfo {author} {\bibfnamefont
  {S.~D.}\ \bibnamefont {Wilson}},\ }\bibfield  {title} {\bibinfo {title}
  {Structural evolution of the kagome superconductors \( \mathrm{AV_3Sb_5} \)
  (\( \mathrm{A=K, Rb,}\) and \(\mathrm{Cs}) \) through charge density wave
  order},\ }\href {https://doi.org/10.1103/PhysRevMaterials.7.024806}
  {\bibfield  {journal} {\bibinfo  {journal} {Physical Review Materials}\
  }\textbf {\bibinfo {volume} {7}},\ \bibinfo {pages} {024806} (\bibinfo {year}
  {2023}{\natexlab{a}})}\BibitemShut {NoStop}%
\bibitem [{\citenamefont {Kang}\ \emph {et~al.}(2023)\citenamefont {Kang},
  \citenamefont {Fang}, \citenamefont {Yoo}, \citenamefont {Ortiz},
  \citenamefont {Oey}, \citenamefont {Choi}, \citenamefont {Ryu}, \citenamefont
  {Kim}, \citenamefont {Jozwiak}, \citenamefont {Bostwick}, \citenamefont
  {Rotenberg}, \citenamefont {Kaxiras}, \citenamefont {Checkelsky},
  \citenamefont {Wilson}, \citenamefont {Park},\ and\ \citenamefont
  {Comin}}]{kang2023charge}%
  \BibitemOpen
  \bibfield  {author} {\bibinfo {author} {\bibfnamefont {M.}~\bibnamefont
  {Kang}}, \bibinfo {author} {\bibfnamefont {S.}~\bibnamefont {Fang}}, \bibinfo
  {author} {\bibfnamefont {J.}~\bibnamefont {Yoo}}, \bibinfo {author}
  {\bibfnamefont {B.~R.}\ \bibnamefont {Ortiz}}, \bibinfo {author}
  {\bibfnamefont {Y.~M.}\ \bibnamefont {Oey}}, \bibinfo {author} {\bibfnamefont
  {J.}~\bibnamefont {Choi}}, \bibinfo {author} {\bibfnamefont {S.~H.}\
  \bibnamefont {Ryu}}, \bibinfo {author} {\bibfnamefont {J.}~\bibnamefont
  {Kim}}, \bibinfo {author} {\bibfnamefont {C.}~\bibnamefont {Jozwiak}},
  \bibinfo {author} {\bibfnamefont {A.}~\bibnamefont {Bostwick}}, \bibinfo
  {author} {\bibfnamefont {E.}~\bibnamefont {Rotenberg}}, \bibinfo {author}
  {\bibfnamefont {E.}~\bibnamefont {Kaxiras}}, \bibinfo {author} {\bibfnamefont
  {J.~G.}\ \bibnamefont {Checkelsky}}, \bibinfo {author} {\bibfnamefont
  {S.~D.}\ \bibnamefont {Wilson}}, \bibinfo {author} {\bibfnamefont {J.-H.}\
  \bibnamefont {Park}},\ and\ \bibinfo {author} {\bibfnamefont
  {R.}~\bibnamefont {Comin}},\ }\bibfield  {title} {\bibinfo {title} {Charge
  order landscape and competition with superconductivity in kagome metals},\
  }\href {https://doi.org/10.1038/s41563-022-01375-2} {\bibfield  {journal}
  {\bibinfo  {journal} {Nature Materials}\ }\textbf {\bibinfo {volume} {22}},\
  \bibinfo {pages} {186} (\bibinfo {year} {2023})}\BibitemShut {NoStop}%
\bibitem [{\citenamefont {Hu}\ \emph {et~al.}(2022)\citenamefont {Hu},
  \citenamefont {Wu}, \citenamefont {Ortiz}, \citenamefont {Han}, \citenamefont
  {Plumb}, \citenamefont {Wilson}, \citenamefont {Schnyder},\ and\
  \citenamefont {Shi}}]{hu2022coexistence}%
  \BibitemOpen
  \bibfield  {author} {\bibinfo {author} {\bibfnamefont {Y.}~\bibnamefont
  {Hu}}, \bibinfo {author} {\bibfnamefont {X.}~\bibnamefont {Wu}}, \bibinfo
  {author} {\bibfnamefont {B.~R.}\ \bibnamefont {Ortiz}}, \bibinfo {author}
  {\bibfnamefont {X.}~\bibnamefont {Han}}, \bibinfo {author} {\bibfnamefont
  {N.~C.}\ \bibnamefont {Plumb}}, \bibinfo {author} {\bibfnamefont {S.~D.}\
  \bibnamefont {Wilson}}, \bibinfo {author} {\bibfnamefont {A.~P.}\
  \bibnamefont {Schnyder}},\ and\ \bibinfo {author} {\bibfnamefont
  {M.}~\bibnamefont {Shi}},\ }\bibfield  {title} {\bibinfo {title} {Coexistence
  of trihexagonal and star-of-david pattern in the charge density wave of the
  kagome superconductor \( \mathrm{AV_3Sb_5} \)},\ }\href
  {https://doi.org/10.1103/PhysRevB.106.L241106} {\bibfield  {journal}
  {\bibinfo  {journal} {Physical Review B}\ }\textbf {\bibinfo {volume}
  {106}},\ \bibinfo {pages} {L241106} (\bibinfo {year} {2022})}\BibitemShut
  {NoStop}%
\bibitem [{\citenamefont {Stahl}\ \emph {et~al.}(2022)\citenamefont {Stahl},
  \citenamefont {Chen}, \citenamefont {Ritschel}, \citenamefont {Shekhar},
  \citenamefont {Sadrollahi}, \citenamefont {Rahn}, \citenamefont {Ivashko},
  \citenamefont {Zimmermann}, \citenamefont {Felser},\ and\ \citenamefont
  {Geck}}]{stahl2022temperature}%
  \BibitemOpen
  \bibfield  {author} {\bibinfo {author} {\bibfnamefont {Q.}~\bibnamefont
  {Stahl}}, \bibinfo {author} {\bibfnamefont {D.}~\bibnamefont {Chen}},
  \bibinfo {author} {\bibfnamefont {T.}~\bibnamefont {Ritschel}}, \bibinfo
  {author} {\bibfnamefont {C.}~\bibnamefont {Shekhar}}, \bibinfo {author}
  {\bibfnamefont {E.}~\bibnamefont {Sadrollahi}}, \bibinfo {author}
  {\bibfnamefont {M.}~\bibnamefont {Rahn}}, \bibinfo {author} {\bibfnamefont
  {O.}~\bibnamefont {Ivashko}}, \bibinfo {author} {\bibfnamefont {M.~v.}\
  \bibnamefont {Zimmermann}}, \bibinfo {author} {\bibfnamefont
  {C.}~\bibnamefont {Felser}},\ and\ \bibinfo {author} {\bibfnamefont
  {J.}~\bibnamefont {Geck}},\ }\bibfield  {title} {\bibinfo {title}
  {Temperature-driven reorganization of electronic order in \(
  \mathrm{CsV_3Sb_5} \)},\ }\href {https://doi.org/10.1103/PhysRevB.105.195136}
  {\bibfield  {journal} {\bibinfo  {journal} {Physical Review B}\ }\textbf
  {\bibinfo {volume} {105}},\ \bibinfo {pages} {195136} (\bibinfo {year}
  {2022})}\BibitemShut {NoStop}%
\bibitem [{\citenamefont {Xiao}\ \emph {et~al.}(2023)\citenamefont {Xiao},
  \citenamefont {Lin}, \citenamefont {Li}, \citenamefont {Zheng}, \citenamefont
  {Francoual}, \citenamefont {Plueckthun}, \citenamefont {Xia}, \citenamefont
  {Qiu}, \citenamefont {Zhang},\ and\ \citenamefont
  {Guo}}]{xiao2023coexistence}%
  \BibitemOpen
  \bibfield  {author} {\bibinfo {author} {\bibfnamefont {Q.}~\bibnamefont
  {Xiao}}, \bibinfo {author} {\bibfnamefont {Y.}~\bibnamefont {Lin}}, \bibinfo
  {author} {\bibfnamefont {Q.}~\bibnamefont {Li}}, \bibinfo {author}
  {\bibfnamefont {X.}~\bibnamefont {Zheng}}, \bibinfo {author} {\bibfnamefont
  {S.}~\bibnamefont {Francoual}}, \bibinfo {author} {\bibfnamefont
  {C.}~\bibnamefont {Plueckthun}}, \bibinfo {author} {\bibfnamefont
  {W.}~\bibnamefont {Xia}}, \bibinfo {author} {\bibfnamefont {Q.}~\bibnamefont
  {Qiu}}, \bibinfo {author} {\bibfnamefont {S.}~\bibnamefont {Zhang}},\ and\
  \bibinfo {author} {\bibfnamefont {Y.}~\bibnamefont {Guo}},\ }\bibfield
  {title} {\bibinfo {title} {Coexistence of multiple stacking charge density
  waves in kagome superconductor \( \mathrm{CsV_3Sb_5} \)},\ }\href
  {https://doi.org/10.1103/PhysRevResearch.5.L012032} {\bibfield  {journal}
  {\bibinfo  {journal} {Physical Review Research}\ }\textbf {\bibinfo {volume}
  {5}},\ \bibinfo {pages} {L012032} (\bibinfo {year} {2023})}\BibitemShut
  {NoStop}%
\bibitem [{\citenamefont {Li}\ \emph {et~al.}(2021)\citenamefont {Li},
  \citenamefont {Zhang}, \citenamefont {Yilmaz}, \citenamefont {Pai},
  \citenamefont {Marvinney}, \citenamefont {Said}, \citenamefont {Yin},
  \citenamefont {Gong}, \citenamefont {Tu}, \citenamefont {Vescovo},
  \citenamefont {Nelson}, \citenamefont {Moore}, \citenamefont {Murakami},
  \citenamefont {Lei}, \citenamefont {Lee}, \citenamefont {Lawrie},\ and\
  \citenamefont {Miao}}]{li2021observation}%
  \BibitemOpen
  \bibfield  {author} {\bibinfo {author} {\bibfnamefont {H.}~\bibnamefont
  {Li}}, \bibinfo {author} {\bibfnamefont {T.~T.}\ \bibnamefont {Zhang}},
  \bibinfo {author} {\bibfnamefont {T.}~\bibnamefont {Yilmaz}}, \bibinfo
  {author} {\bibfnamefont {Y.~Y.}\ \bibnamefont {Pai}}, \bibinfo {author}
  {\bibfnamefont {C.~E.}\ \bibnamefont {Marvinney}}, \bibinfo {author}
  {\bibfnamefont {A.}~\bibnamefont {Said}}, \bibinfo {author} {\bibfnamefont
  {Q.~W.}\ \bibnamefont {Yin}}, \bibinfo {author} {\bibfnamefont {C.~S.}\
  \bibnamefont {Gong}}, \bibinfo {author} {\bibfnamefont {Z.~J.}\ \bibnamefont
  {Tu}}, \bibinfo {author} {\bibfnamefont {E.}~\bibnamefont {Vescovo}},
  \bibinfo {author} {\bibfnamefont {C.~S.}\ \bibnamefont {Nelson}}, \bibinfo
  {author} {\bibfnamefont {R.~G.}\ \bibnamefont {Moore}}, \bibinfo {author}
  {\bibfnamefont {S.}~\bibnamefont {Murakami}}, \bibinfo {author}
  {\bibfnamefont {H.~C.}\ \bibnamefont {Lei}}, \bibinfo {author} {\bibfnamefont
  {H.~N.}\ \bibnamefont {Lee}}, \bibinfo {author} {\bibfnamefont {B.~J.}\
  \bibnamefont {Lawrie}},\ and\ \bibinfo {author} {\bibfnamefont
  {H.}~\bibnamefont {Miao}},\ }\bibfield  {title} {\bibinfo {title}
  {Observation of unconventional charge density wave without acoustic phonon
  anomaly in kagome superconductors ${A\mathrm{V}}_{3}{\mathrm{sb}}_{5}$
  ($a=\mathrm{Rb}$, cs)},\ }\href {https://doi.org/10.1103/PhysRevX.11.031050}
  {\bibfield  {journal} {\bibinfo  {journal} {Phys. Rev. X}\ }\textbf {\bibinfo
  {volume} {11}},\ \bibinfo {pages} {031050} (\bibinfo {year}
  {2021})}\BibitemShut {NoStop}%
\bibitem [{\citenamefont {Ortiz}\ \emph
  {et~al.}(2021{\natexlab{b}})\citenamefont {Ortiz}, \citenamefont {Teicher},
  \citenamefont {Kautzsch}, \citenamefont {Sarte}, \citenamefont {Ratcliff},
  \citenamefont {Harter}, \citenamefont {Ruff}, \citenamefont {Seshadri},\ and\
  \citenamefont {Wilson}}]{ortiz2021fermi}%
  \BibitemOpen
  \bibfield  {author} {\bibinfo {author} {\bibfnamefont {B.~R.}\ \bibnamefont
  {Ortiz}}, \bibinfo {author} {\bibfnamefont {S.~M.~L.}\ \bibnamefont
  {Teicher}}, \bibinfo {author} {\bibfnamefont {L.}~\bibnamefont {Kautzsch}},
  \bibinfo {author} {\bibfnamefont {P.~M.}\ \bibnamefont {Sarte}}, \bibinfo
  {author} {\bibfnamefont {N.}~\bibnamefont {Ratcliff}}, \bibinfo {author}
  {\bibfnamefont {J.}~\bibnamefont {Harter}}, \bibinfo {author} {\bibfnamefont
  {J.~P.~C.}\ \bibnamefont {Ruff}}, \bibinfo {author} {\bibfnamefont
  {R.}~\bibnamefont {Seshadri}},\ and\ \bibinfo {author} {\bibfnamefont
  {S.~D.}\ \bibnamefont {Wilson}},\ }\bibfield  {title} {\bibinfo {title}
  {Fermi surface mapping and the nature of charge-density-wave order in the
  kagome superconductor \( \mathrm{CsV_3Sb_5} \)},\ }\href
  {https://doi.org/10.1103/PhysRevX.11.041030} {\bibfield  {journal} {\bibinfo
  {journal} {Physical Review X}\ }\textbf {\bibinfo {volume} {11}},\ \bibinfo
  {pages} {041030} (\bibinfo {year} {2021}{\natexlab{b}})}\BibitemShut
  {NoStop}%
\bibitem [{\citenamefont {Wu}\ \emph {et~al.}(2022)\citenamefont {Wu},
  \citenamefont {Ortiz}, \citenamefont {Tan}, \citenamefont {Wilson},
  \citenamefont {Yan}, \citenamefont {Birol},\ and\ \citenamefont
  {Blumberg}}]{wu2022charge}%
  \BibitemOpen
  \bibfield  {author} {\bibinfo {author} {\bibfnamefont {S.}~\bibnamefont
  {Wu}}, \bibinfo {author} {\bibfnamefont {B.~R.}\ \bibnamefont {Ortiz}},
  \bibinfo {author} {\bibfnamefont {H.}~\bibnamefont {Tan}}, \bibinfo {author}
  {\bibfnamefont {S.~D.}\ \bibnamefont {Wilson}}, \bibinfo {author}
  {\bibfnamefont {B.}~\bibnamefont {Yan}}, \bibinfo {author} {\bibfnamefont
  {T.}~\bibnamefont {Birol}},\ and\ \bibinfo {author} {\bibfnamefont
  {G.}~\bibnamefont {Blumberg}},\ }\bibfield  {title} {\bibinfo {title} {Charge
  density wave order in the kagome metal \( \mathrm{AV_3Sb_5} \) (\(
  \mathrm{A=Cs,Rb, K} \))},\ }\href
  {https://doi.org/10.1103/PhysRevB.105.155106} {\bibfield  {journal} {\bibinfo
   {journal} {Physical Review B}\ }\textbf {\bibinfo {volume} {105}},\ \bibinfo
  {pages} {155106} (\bibinfo {year} {2022})}\BibitemShut {NoStop}%
\bibitem [{\citenamefont {Li}\ \emph {et~al.}(2022)\citenamefont {Li},
  \citenamefont {Zhao}, \citenamefont {Ortiz}, \citenamefont {Park},
  \citenamefont {Ye}, \citenamefont {Balents}, \citenamefont {Wang},
  \citenamefont {Wilson},\ and\ \citenamefont {Zeljkovic}}]{li2022rotation}%
  \BibitemOpen
  \bibfield  {author} {\bibinfo {author} {\bibfnamefont {H.}~\bibnamefont
  {Li}}, \bibinfo {author} {\bibfnamefont {H.}~\bibnamefont {Zhao}}, \bibinfo
  {author} {\bibfnamefont {B.~R.}\ \bibnamefont {Ortiz}}, \bibinfo {author}
  {\bibfnamefont {T.}~\bibnamefont {Park}}, \bibinfo {author} {\bibfnamefont
  {M.}~\bibnamefont {Ye}}, \bibinfo {author} {\bibfnamefont {L.}~\bibnamefont
  {Balents}}, \bibinfo {author} {\bibfnamefont {Z.}~\bibnamefont {Wang}},
  \bibinfo {author} {\bibfnamefont {S.~D.}\ \bibnamefont {Wilson}},\ and\
  \bibinfo {author} {\bibfnamefont {I.}~\bibnamefont {Zeljkovic}},\ }\bibfield
  {title} {\bibinfo {title} {Rotation symmetry breaking in the normal state of
  a kagome superconductor \( \mathrm{KV_3Sb_5} \)},\ }\href
  {https://doi.org/10.1038/s41567-021-01479-7} {\bibfield  {journal} {\bibinfo
  {journal} {Nature Physics}\ }\textbf {\bibinfo {volume} {18}},\ \bibinfo
  {pages} {265} (\bibinfo {year} {2022})}\BibitemShut {NoStop}%
\bibitem [{\citenamefont {Osborne}(2024)}]{osbornenexpy}%
  \BibitemOpen
  \bibfield  {author} {\bibinfo {author} {\bibfnamefont {R.}~\bibnamefont
  {Osborne}},\ }\href {https://nexpy.github.io/nexpy/} {\bibinfo {title}
  {Nexpy: A python gui to analyze nexus data}} (\bibinfo {year}
  {2024})\BibitemShut {NoStop}%
\bibitem [{\citenamefont {Bruker}(2024)}]{APEX3}%
  \BibitemOpen
  \bibfield  {author} {\bibinfo {author} {\bibnamefont {Bruker}},\ }\href
  {https://www.brukersupport.com/ProductDetail/3177} {\bibinfo {title} {Apex3:
  Software for chemical crystallography}} (\bibinfo {year} {2024})\BibitemShut
  {NoStop}%
\bibitem [{\citenamefont {Venderley}\ \emph {et~al.}(2022)\citenamefont
  {Venderley}, \citenamefont {Mallayya}, \citenamefont {Matty}, \citenamefont
  {Krogstad}, \citenamefont {Ruff}, \citenamefont {Pleiss}, \citenamefont
  {Kishore}, \citenamefont {Mandrus}, \citenamefont {Phelan}, \citenamefont
  {Poudel}, \citenamefont {Wilson}, \citenamefont {Weinberger}, \citenamefont
  {Upreti}, \citenamefont {Norman}, \citenamefont {Rosenkranz}, \citenamefont
  {Osborn},\ and\ \citenamefont {Kim}}]{Venderley2022Proc.Natl.Acad.Sci.}%
  \BibitemOpen
  \bibfield  {author} {\bibinfo {author} {\bibfnamefont {J.}~\bibnamefont
  {Venderley}}, \bibinfo {author} {\bibfnamefont {K.}~\bibnamefont {Mallayya}},
  \bibinfo {author} {\bibfnamefont {M.}~\bibnamefont {Matty}}, \bibinfo
  {author} {\bibfnamefont {M.}~\bibnamefont {Krogstad}}, \bibinfo {author}
  {\bibfnamefont {J.}~\bibnamefont {Ruff}}, \bibinfo {author} {\bibfnamefont
  {G.}~\bibnamefont {Pleiss}}, \bibinfo {author} {\bibfnamefont
  {V.}~\bibnamefont {Kishore}}, \bibinfo {author} {\bibfnamefont
  {D.}~\bibnamefont {Mandrus}}, \bibinfo {author} {\bibfnamefont
  {D.}~\bibnamefont {Phelan}}, \bibinfo {author} {\bibfnamefont
  {L.}~\bibnamefont {Poudel}}, \bibinfo {author} {\bibfnamefont {A.~G.}\
  \bibnamefont {Wilson}}, \bibinfo {author} {\bibfnamefont {K.}~\bibnamefont
  {Weinberger}}, \bibinfo {author} {\bibfnamefont {P.}~\bibnamefont {Upreti}},
  \bibinfo {author} {\bibfnamefont {M.}~\bibnamefont {Norman}}, \bibinfo
  {author} {\bibfnamefont {S.}~\bibnamefont {Rosenkranz}}, \bibinfo {author}
  {\bibfnamefont {R.}~\bibnamefont {Osborn}},\ and\ \bibinfo {author}
  {\bibfnamefont {E.-A.}\ \bibnamefont {Kim}},\ }\bibfield  {title} {\bibinfo
  {title} {Harnessing interpretable and unsupervised machine learning to
  address big data from modern {{X-ray}} diffraction},\ }\href
  {https://doi.org/10.1073/pnas.2109665119} {\bibfield  {journal} {\bibinfo
  {journal} {Proceedings of the National Academy of Sciences}\ }\textbf
  {\bibinfo {volume} {119}},\ \bibinfo {pages} {e2109665119} (\bibinfo {year}
  {2022})}\BibitemShut {NoStop}%
\bibitem [{\citenamefont {Simons}\ \emph {et~al.}(2016)\citenamefont {Simons},
  \citenamefont {Jakobsen}, \citenamefont {Ahl}, \citenamefont {Detlefs},\ and\
  \citenamefont {Poulsen}}]{simons2016multiscale}%
  \BibitemOpen
  \bibfield  {author} {\bibinfo {author} {\bibfnamefont {H.}~\bibnamefont
  {Simons}}, \bibinfo {author} {\bibfnamefont {A.~C.}\ \bibnamefont
  {Jakobsen}}, \bibinfo {author} {\bibfnamefont {S.~R.}\ \bibnamefont {Ahl}},
  \bibinfo {author} {\bibfnamefont {C.}~\bibnamefont {Detlefs}},\ and\ \bibinfo
  {author} {\bibfnamefont {H.~F.}\ \bibnamefont {Poulsen}},\ }\bibfield
  {title} {\bibinfo {title} {Multiscale 3d characterization with dark-field
  x-ray microscopy},\ }\href {https://doi.org/10.1557/mrs.2016.114} {\bibfield
  {journal} {\bibinfo  {journal} {Mrs Bulletin}\ }\textbf {\bibinfo {volume}
  {41}},\ \bibinfo {pages} {454} (\bibinfo {year} {2016})}\BibitemShut
  {NoStop}%
\bibitem [{\citenamefont {Yildirim}\ \emph {et~al.}(2020)\citenamefont
  {Yildirim}, \citenamefont {Cook}, \citenamefont {Detlefs}, \citenamefont
  {Simons},\ and\ \citenamefont {Poulsen}}]{yildirim2020probing}%
  \BibitemOpen
  \bibfield  {author} {\bibinfo {author} {\bibfnamefont {C.}~\bibnamefont
  {Yildirim}}, \bibinfo {author} {\bibfnamefont {P.}~\bibnamefont {Cook}},
  \bibinfo {author} {\bibfnamefont {C.}~\bibnamefont {Detlefs}}, \bibinfo
  {author} {\bibfnamefont {H.}~\bibnamefont {Simons}},\ and\ \bibinfo {author}
  {\bibfnamefont {H.~F.}\ \bibnamefont {Poulsen}},\ }\bibfield  {title}
  {\bibinfo {title} {Probing nanoscale structure and strain by dark-field x-ray
  microscopy},\ }\href {https://doi.org/10.1557/mrs.2020.89} {\bibfield
  {journal} {\bibinfo  {journal} {MRS Bulletin}\ }\textbf {\bibinfo {volume}
  {45}},\ \bibinfo {pages} {277} (\bibinfo {year} {2020})}\BibitemShut
  {NoStop}%
\bibitem [{\citenamefont {Plumb}\ \emph {et~al.}(2023)\citenamefont {Plumb},
  \citenamefont {Poudyal}, \citenamefont {Dally}, \citenamefont {Daly},
  \citenamefont {Wilson},\ and\ \citenamefont {Islam}}]{plumb2023dark}%
  \BibitemOpen
  \bibfield  {author} {\bibinfo {author} {\bibfnamefont {J.}~\bibnamefont
  {Plumb}}, \bibinfo {author} {\bibfnamefont {I.}~\bibnamefont {Poudyal}},
  \bibinfo {author} {\bibfnamefont {R.~L.}\ \bibnamefont {Dally}}, \bibinfo
  {author} {\bibfnamefont {S.}~\bibnamefont {Daly}}, \bibinfo {author}
  {\bibfnamefont {S.~D.}\ \bibnamefont {Wilson}},\ and\ \bibinfo {author}
  {\bibfnamefont {Z.}~\bibnamefont {Islam}},\ }\bibfield  {title} {\bibinfo
  {title} {Dark field x-ray microscopy below liquid-helium temperature: The
  case of \( \mathrm{NaMnO_2} \)},\ }\href
  {https://doi.org/10.1016/j.matchar.2023.113174} {\bibfield  {journal}
  {\bibinfo  {journal} {Materials Characterization}\ }\textbf {\bibinfo
  {volume} {204}},\ \bibinfo {pages} {113174} (\bibinfo {year}
  {2023})}\BibitemShut {NoStop}%
\bibitem [{\citenamefont {Opolka}\ \emph {et~al.}(2021)\citenamefont {Opolka},
  \citenamefont {M{\"u}ller}, \citenamefont {Fella}, \citenamefont {Balles},
  \citenamefont {Mohr},\ and\ \citenamefont {Last}}]{opolka2021multi}%
  \BibitemOpen
  \bibfield  {author} {\bibinfo {author} {\bibfnamefont {A.}~\bibnamefont
  {Opolka}}, \bibinfo {author} {\bibfnamefont {D.}~\bibnamefont {M{\"u}ller}},
  \bibinfo {author} {\bibfnamefont {C.}~\bibnamefont {Fella}}, \bibinfo
  {author} {\bibfnamefont {A.}~\bibnamefont {Balles}}, \bibinfo {author}
  {\bibfnamefont {J.}~\bibnamefont {Mohr}},\ and\ \bibinfo {author}
  {\bibfnamefont {A.}~\bibnamefont {Last}},\ }\bibfield  {title} {\bibinfo
  {title} {Multi-lens array full-field x-ray microscopy},\ }\href
  {https://doi.org/10.3390/app11167234} {\bibfield  {journal} {\bibinfo
  {journal} {Applied Sciences}\ }\textbf {\bibinfo {volume} {11}},\ \bibinfo
  {pages} {7234} (\bibinfo {year} {2021})}\BibitemShut {NoStop}%
\bibitem [{\citenamefont {Qiao}\ \emph {et~al.}(2020)\citenamefont {Qiao},
  \citenamefont {Shi}, \citenamefont {Kenesei}, \citenamefont {Last},
  \citenamefont {Assoufid},\ and\ \citenamefont {Islam}}]{qiao2020large}%
  \BibitemOpen
  \bibfield  {author} {\bibinfo {author} {\bibfnamefont {Z.}~\bibnamefont
  {Qiao}}, \bibinfo {author} {\bibfnamefont {X.}~\bibnamefont {Shi}}, \bibinfo
  {author} {\bibfnamefont {P.}~\bibnamefont {Kenesei}}, \bibinfo {author}
  {\bibfnamefont {A.}~\bibnamefont {Last}}, \bibinfo {author} {\bibfnamefont
  {L.}~\bibnamefont {Assoufid}},\ and\ \bibinfo {author} {\bibfnamefont
  {Z.}~\bibnamefont {Islam}},\ }\bibfield  {title} {\bibinfo {title} {A large
  field-of-view high-resolution hard x-ray microscope using polymer optics},\
  }\bibfield  {journal} {\bibinfo  {journal} {Review of Scientific
  Instruments}\ }\textbf {\bibinfo {volume} {91}},\ \href
  {https://doi.org/10.1063/5.0011961} {10.1063/5.0011961} (\bibinfo {year}
  {2020})\BibitemShut {NoStop}%
\bibitem [{\citenamefont {G{\"u}rsoy}\ \emph {et~al.}(2024)\citenamefont
  {G{\"u}rsoy}, \citenamefont {Yay}, \citenamefont {Kisiel}, \citenamefont
  {Wojcik}, \citenamefont {Sheyfer}, \citenamefont {Highland}, \citenamefont
  {Fisher}, \citenamefont {Hruszkewycz},\ and\ \citenamefont
  {Islam}}]{gursoy2024dark}%
  \BibitemOpen
  \bibfield  {author} {\bibinfo {author} {\bibfnamefont {D.}~\bibnamefont
  {G{\"u}rsoy}}, \bibinfo {author} {\bibfnamefont {K.~A.}\ \bibnamefont {Yay}},
  \bibinfo {author} {\bibfnamefont {E.}~\bibnamefont {Kisiel}}, \bibinfo
  {author} {\bibfnamefont {M.}~\bibnamefont {Wojcik}}, \bibinfo {author}
  {\bibfnamefont {D.}~\bibnamefont {Sheyfer}}, \bibinfo {author} {\bibfnamefont
  {M.}~\bibnamefont {Highland}}, \bibinfo {author} {\bibfnamefont {I.~R.}\
  \bibnamefont {Fisher}}, \bibinfo {author} {\bibfnamefont {S.}~\bibnamefont
  {Hruszkewycz}},\ and\ \bibinfo {author} {\bibfnamefont {Z.}~\bibnamefont
  {Islam}},\ }\bibfield  {title} {\bibinfo {title} {Dark-field x-ray microscopy
  with structured illumination for three-dimensional imaging},\ }\href
  {https://arxiv.org/abs/2405.12799} {\bibfield  {journal} {\bibinfo  {journal}
  {arXiv preprint arXiv:2405.12799}\ } (\bibinfo {year} {2024})}\BibitemShut
  {NoStop}%
\bibitem [{\citenamefont {Garriga~Ferrer}\ \emph {et~al.}(2023)\citenamefont
  {Garriga~Ferrer}, \citenamefont {Rodr{\'\i}guez-Lamas}, \citenamefont
  {Payno}, \citenamefont {De~Nolf}, \citenamefont {Cook}, \citenamefont
  {Sol{\'e}~Jover}, \citenamefont {Yildirim},\ and\ \citenamefont
  {Detlefs}}]{garriga2023darfix}%
  \BibitemOpen
  \bibfield  {author} {\bibinfo {author} {\bibfnamefont {J.}~\bibnamefont
  {Garriga~Ferrer}}, \bibinfo {author} {\bibfnamefont {R.}~\bibnamefont
  {Rodr{\'\i}guez-Lamas}}, \bibinfo {author} {\bibfnamefont {H.}~\bibnamefont
  {Payno}}, \bibinfo {author} {\bibfnamefont {W.}~\bibnamefont {De~Nolf}},
  \bibinfo {author} {\bibfnamefont {P.}~\bibnamefont {Cook}}, \bibinfo {author}
  {\bibfnamefont {V.~A.}\ \bibnamefont {Sol{\'e}~Jover}}, \bibinfo {author}
  {\bibfnamefont {C.}~\bibnamefont {Yildirim}},\ and\ \bibinfo {author}
  {\bibfnamefont {C.}~\bibnamefont {Detlefs}},\ }\bibfield  {title} {\bibinfo
  {title} {darfix--data analysis for dark-field x-ray microscopy},\ }\bibfield
  {journal} {\bibinfo  {journal} {Journal of Synchrotron Radiation}\ }\textbf
  {\bibinfo {volume} {30}},\ \href {https://doi.org/10.1107/S1600577523001674}
  {10.1107/S1600577523001674} (\bibinfo {year} {2023})\BibitemShut {NoStop}%
\bibitem [{\citenamefont {Kautzsch}\ \emph
  {et~al.}(2023{\natexlab{b}})\citenamefont {Kautzsch}, \citenamefont {Oey},
  \citenamefont {Li}, \citenamefont {Ren}, \citenamefont {Ortiz}, \citenamefont
  {Pokharel}, \citenamefont {Seshadri}, \citenamefont {Ruff}, \citenamefont
  {Kongruengkit}, \citenamefont {Harter}, \citenamefont {Wang}, \citenamefont
  {Zeljkovic},\ and\ \citenamefont {Wilson}}]{Kautzsch2023}%
  \BibitemOpen
  \bibfield  {author} {\bibinfo {author} {\bibfnamefont {L.}~\bibnamefont
  {Kautzsch}}, \bibinfo {author} {\bibfnamefont {Y.~M.}\ \bibnamefont {Oey}},
  \bibinfo {author} {\bibfnamefont {H.}~\bibnamefont {Li}}, \bibinfo {author}
  {\bibfnamefont {Z.}~\bibnamefont {Ren}}, \bibinfo {author} {\bibfnamefont
  {B.~R.}\ \bibnamefont {Ortiz}}, \bibinfo {author} {\bibfnamefont
  {G.}~\bibnamefont {Pokharel}}, \bibinfo {author} {\bibfnamefont
  {R.}~\bibnamefont {Seshadri}}, \bibinfo {author} {\bibfnamefont
  {J.}~\bibnamefont {Ruff}}, \bibinfo {author} {\bibfnamefont {T.}~\bibnamefont
  {Kongruengkit}}, \bibinfo {author} {\bibfnamefont {J.~W.}\ \bibnamefont
  {Harter}}, \bibinfo {author} {\bibfnamefont {Z.}~\bibnamefont {Wang}},
  \bibinfo {author} {\bibfnamefont {I.}~\bibnamefont {Zeljkovic}},\ and\
  \bibinfo {author} {\bibfnamefont {S.~D.}\ \bibnamefont {Wilson}},\ }\bibfield
   {title} {\bibinfo {title} {Incommensurate charge-stripe correlations in the
  kagome superconductor \( \mathrm{CsV_3Sb_{5-x}Sn_x} \)},\ }\href
  {https://doi.org/10.1038/s41535-023-00570-x} {\bibfield  {journal} {\bibinfo
  {journal} {npj Quantum Materials}\ }\textbf {\bibinfo {volume} {8}},\
  \bibinfo {pages} {37} (\bibinfo {year} {2023}{\natexlab{b}})}\BibitemShut
  {NoStop}%
\bibitem [{\citenamefont {Zheng}\ \emph {et~al.}(2022)\citenamefont {Zheng},
  \citenamefont {Wu}, \citenamefont {Yang}, \citenamefont {Nie}, \citenamefont
  {Shan}, \citenamefont {Sun}, \citenamefont {Song}, \citenamefont {Yu},
  \citenamefont {Li}, \citenamefont {Zhao}, \citenamefont {Li}, \citenamefont
  {Kang}, \citenamefont {Zhou}, \citenamefont {Liu}, \citenamefont {Xiang},
  \citenamefont {Ying}, \citenamefont {Wang}, \citenamefont {Wu},\ and\
  \citenamefont {Chen}}]{Zheng2022}%
  \BibitemOpen
  \bibfield  {author} {\bibinfo {author} {\bibfnamefont {L.}~\bibnamefont
  {Zheng}}, \bibinfo {author} {\bibfnamefont {Z.}~\bibnamefont {Wu}}, \bibinfo
  {author} {\bibfnamefont {Y.}~\bibnamefont {Yang}}, \bibinfo {author}
  {\bibfnamefont {L.}~\bibnamefont {Nie}}, \bibinfo {author} {\bibfnamefont
  {M.}~\bibnamefont {Shan}}, \bibinfo {author} {\bibfnamefont {K.}~\bibnamefont
  {Sun}}, \bibinfo {author} {\bibfnamefont {D.}~\bibnamefont {Song}}, \bibinfo
  {author} {\bibfnamefont {F.}~\bibnamefont {Yu}}, \bibinfo {author}
  {\bibfnamefont {J.}~\bibnamefont {Li}}, \bibinfo {author} {\bibfnamefont
  {D.}~\bibnamefont {Zhao}}, \bibinfo {author} {\bibfnamefont {S.}~\bibnamefont
  {Li}}, \bibinfo {author} {\bibfnamefont {B.}~\bibnamefont {Kang}}, \bibinfo
  {author} {\bibfnamefont {Y.}~\bibnamefont {Zhou}}, \bibinfo {author}
  {\bibfnamefont {K.}~\bibnamefont {Liu}}, \bibinfo {author} {\bibfnamefont
  {Z.}~\bibnamefont {Xiang}}, \bibinfo {author} {\bibfnamefont
  {J.}~\bibnamefont {Ying}}, \bibinfo {author} {\bibfnamefont {Z.}~\bibnamefont
  {Wang}}, \bibinfo {author} {\bibfnamefont {T.}~\bibnamefont {Wu}},\ and\
  \bibinfo {author} {\bibfnamefont {X.}~\bibnamefont {Chen}},\ }\bibfield
  {title} {\bibinfo {title} {Emergent charge order in pressurized kagome
  superconductor \( \mathrm{AV_3Sb_5} \)},\ }\href
  {https://doi.org/10.1038/s41586-022-05351-3} {\bibfield  {journal} {\bibinfo
  {journal} {Nature}\ }\textbf {\bibinfo {volume} {611}},\ \bibinfo {pages}
  {682} (\bibinfo {year} {2022})}\BibitemShut {NoStop}%
\bibitem [{\citenamefont {Guo}\ \emph {et~al.}(2022)\citenamefont {Guo},
  \citenamefont {Putzke}, \citenamefont {Konyzheva}, \citenamefont {Huang},
  \citenamefont {Gutierrez-Amigo}, \citenamefont {Errea}, \citenamefont {Chen},
  \citenamefont {Vergniory}, \citenamefont {Felser}, \citenamefont {Fischer},
  \citenamefont {Neupert},\ and\ \citenamefont {Moll}}]{guo2022switchable}%
  \BibitemOpen
  \bibfield  {author} {\bibinfo {author} {\bibfnamefont {C.}~\bibnamefont
  {Guo}}, \bibinfo {author} {\bibfnamefont {C.}~\bibnamefont {Putzke}},
  \bibinfo {author} {\bibfnamefont {S.}~\bibnamefont {Konyzheva}}, \bibinfo
  {author} {\bibfnamefont {X.}~\bibnamefont {Huang}}, \bibinfo {author}
  {\bibfnamefont {M.}~\bibnamefont {Gutierrez-Amigo}}, \bibinfo {author}
  {\bibfnamefont {I.}~\bibnamefont {Errea}}, \bibinfo {author} {\bibfnamefont
  {D.}~\bibnamefont {Chen}}, \bibinfo {author} {\bibfnamefont {M.~G.}\
  \bibnamefont {Vergniory}}, \bibinfo {author} {\bibfnamefont {C.}~\bibnamefont
  {Felser}}, \bibinfo {author} {\bibfnamefont {M.~H.}\ \bibnamefont {Fischer}},
  \bibinfo {author} {\bibfnamefont {T.}~\bibnamefont {Neupert}},\ and\ \bibinfo
  {author} {\bibfnamefont {P.~J.~W.}\ \bibnamefont {Moll}},\ }\bibfield
  {title} {\bibinfo {title} {Switchable chiral transport in charge-ordered
  kagome metal \( \mathrm{CsV_3Sb_5} \)},\ }\href
  {https://doi.org/10.1038/s41586-022-05127-9} {\bibfield  {journal} {\bibinfo
  {journal} {Nature}\ }\textbf {\bibinfo {volume} {611}},\ \bibinfo {pages}
  {461} (\bibinfo {year} {2022})}\BibitemShut {NoStop}%
\bibitem [{\citenamefont {Guo}\ \emph {et~al.}(2024)\citenamefont {Guo},
  \citenamefont {Wagner}, \citenamefont {Putzke}, \citenamefont {Chen},
  \citenamefont {Wang}, \citenamefont {Zhang}, \citenamefont {Gutierrez-Amigo},
  \citenamefont {Errea}, \citenamefont {Vergniory}, \citenamefont {Felser},
  \citenamefont {Fischer}, \citenamefont {Neupert},\ and\ \citenamefont
  {Moll}}]{guo2024correlated}%
  \BibitemOpen
  \bibfield  {author} {\bibinfo {author} {\bibfnamefont {C.}~\bibnamefont
  {Guo}}, \bibinfo {author} {\bibfnamefont {G.}~\bibnamefont {Wagner}},
  \bibinfo {author} {\bibfnamefont {C.}~\bibnamefont {Putzke}}, \bibinfo
  {author} {\bibfnamefont {D.}~\bibnamefont {Chen}}, \bibinfo {author}
  {\bibfnamefont {K.}~\bibnamefont {Wang}}, \bibinfo {author} {\bibfnamefont
  {L.}~\bibnamefont {Zhang}}, \bibinfo {author} {\bibfnamefont
  {M.}~\bibnamefont {Gutierrez-Amigo}}, \bibinfo {author} {\bibfnamefont
  {I.}~\bibnamefont {Errea}}, \bibinfo {author} {\bibfnamefont {M.~G.}\
  \bibnamefont {Vergniory}}, \bibinfo {author} {\bibfnamefont {C.}~\bibnamefont
  {Felser}}, \bibinfo {author} {\bibfnamefont {M.~H.}\ \bibnamefont {Fischer}},
  \bibinfo {author} {\bibfnamefont {T.}~\bibnamefont {Neupert}},\ and\ \bibinfo
  {author} {\bibfnamefont {P.~J.~W.}\ \bibnamefont {Moll}},\ }\bibfield
  {title} {\bibinfo {title} {Correlated order at the tipping point in the
  kagome metal \( \mathrm{CsV_3Sb_5} \)},\ }\href
  {https://doi.org/10.1038/s41567-023-02374-z} {\bibfield  {journal} {\bibinfo
  {journal} {Nature Physics}\ }\textbf {\bibinfo {volume} {20}},\ \bibinfo
  {pages} {579} (\bibinfo {year} {2024})}\BibitemShut {NoStop}%
\end{thebibliography}%

\end{document}